\documentclass[a4paper,11pt]{article}
\usepackage{jheppub} 

\usepackage[utf8]{inputenc}
\usepackage{mathrsfs}
\usepackage{amssymb}	
\usepackage{amsmath}
\usepackage{graphicx}
\usepackage{subdepth}
\usepackage{physics}
\usepackage[x11names]{xcolor}
\usepackage{siunitx}

\usepackage{booktabs}

\usepackage{overpic}

\usepackage{simpler-wick}
\usepackage{slashed}

\usepackage{subcaption}

\newcommand{\ut}{{\mathbf{u}}}

\newcommand{\xt}{{\mathbf{x}}}
\newcommand{\yt}{{\mathbf{y}}}
\newcommand{\zt}{{\mathbf{z}}}
\newcommand{\nc}{N_c}

\newcommand{\as}{\alpha_s}
\newcommand{\gs}{g_s}

\newcommand{\w}{w}

\newcommand{\vol}[1]{\text{vol}\qty(#1)}
\newcommand{\su}{\text{SU}}
\newcommand{\sua}{\mathfrak{su}}
\newcommand{\ua}{\mathfrak{u}}
\newcommand{\sym}{\text{sym}}

\newcommand{\rhosq}{{\vec \rho}^{\, 2}}
\newcommand{\sigmasq}{{\vec \sigma}^{\, 2}}

\newcommand{\DD}[2][]{\Delta^{#1} #2 \,}
\newcommand{\dcal}[1]{\mathcal{D} #1 \,}

\newcommand{\update}[1]{#1}

\title{Color-glass condensate beyond the Gaussian approximation}

\author[a,b]{Jani Penttala}
\affiliation[a]{
Department of Physics and Astronomy, University of California, Los Angeles, CA 90095, USA
}
\affiliation[b]{Mani L. Bhaumik Institute for Theoretical Physics, University of California, Los Angeles, California 90095, USA}
\emailAdd{janipenttala@physics.ucla.edu}

\abstract{
In the high-energy limit, perturbative calculations in QCD are conveniently done using the dipole picture
which factorizes the scattering amplitude into a perturbative part and the nonperturbative scattering off the nuclear target, described using correlators of Wilson lines.
These correlators can be computed in the color-glass condensate effective field theory by using a Gaussian model for the color density of the target.
In this work, we generalize the Gaussian model to a generic function that is local in the \update{light-cone coordinates}, and
show how to compute physical Wilson-line correlators in this model.
We also consider a simple model for the color density based on stable probability distributions and show that the small-dipole behavior of the dipole amplitude is modified from quadratic to a power law, where the power is given by the stability parameter of the distribution.
This generalization of the Gaussian model 
is suitable for numerical applications in the high-energy limit and
can be used in future phenomenological studies of the nuclear structure.
}

\begin{document}

\maketitle
\flushbottom

\section{Introduction}

In high-energy collisions with initial hadrons, 
the gluon component of the target nucleus has been observed to dominate over quarks and antiquarks~\cite{H1:2015ubc}.
This has been understood to follow from the increasing probability of gluon emissions with a small longitudinal momentum component, typically denoted by $x$.
Due to this high occupation number of gluons, the target allows for a description in terms of classical color fields, also called the color-glass condensate~\cite{Iancu:2003xm,Gelis:2010nm}.
Computing observables in this description requires knowledge about the color-field configuration of the target, which is nonperturbative and typically modeled using the McLerran--Venugopalan (MV) model~\cite{McLerran:1993ka,McLerran:1993ni,McLerran:1994vd}  where a given color-field configuration has a probability given by a Gaussian weight function.
This model has achieved widespread success across a broad range of applications, from modeling the target in dilute--dense collisions --- such as deep inelastic scattering and proton--nucleus ($pA$) collisions --- to dense--dense collisions, including heavy-ion collisions where the MV model serves as an initial condition for the formation of quark--gluon plasma~\cite{Gelis:2010nm}.

Several modifications to the standard MV model have also been considered since its inception, such as the inclusion of the odderon contribution~\cite{Kovchegov:2003dm,Jeon:2005cf,Hatta:2005as,Kovchegov:2012ga,Kovner:2016jfp}, subleading correlations~\cite{Jeon:2004rk,Dumitru:2011ax,Dumitru:2011zz,Giannini:2020xme,Giannini:2021ibn},
and helicity dependence~\cite{Cougoulic:2020tbc}.
Additionally, the MV model can be extended from its original local form to non-local correlations between different transverse coordinates~\cite{Blaizot:2004wv,Dominguez:2008aa,Dominguez:2011wm,Lappi:2015vta,Lappi:2017skr}.
This non-local Gaussian form enables the computation of higher-order correlations starting from a model for the dipole amplitude, making the Gaussian approximation a powerful tool in studying more complicated observables.

Along with the original MV model, all of these generalizations use the Gaussian approximation for the color-field correlations in some form.
This is a very natural approximation for two reasons:
First, it has the physical motivation following from the central limit theorem being applied to a large number of random color sources.
Second, it allows analytical expressions of Wilson-line correlators that would otherwise be extremely difficult to compute.
However, while these motivations are very reasonable from the point of theoretical computations, it is also worthwhile to explore the possibility of models beyond the Gaussian form.

In this work, we wish to study a generalized version of classical color-field configurations that relaxes the assumption about the Gaussian weight.
The main motivation is to have more flexibility on the model for the target configuration and study how the general behavior of Wilson-line correlators is changed. 
With the increasing amount of experimental data becoming available in the high-energy regime, it is expected that comparisons to the data should be able to restrict the color-field structure of the target very precisely.

The article is organized as follows.
In Sec.~\ref{sec:cgc}, we briefly summarize the color-glass condensate effective field theory and discuss the approximations leading to the MV model.
We also present the weight function for a  generic class of models considered in this work.
The computation of Wilson-line correlators using this weight function is detailed in Sec.~\ref{sec:wilson-line_correlators}, where we also discuss modeling a heavy nucleus as a collection of nucleons with a given color configuration.
In Sec.~\ref{sec:stable}, we will restrict the class of models to weight functions inspired by stable distributions which are a natural extension of the Gaussian probability density.
Finally, we summarize our results in Sec.~\ref{sec:discussion}.

\section{Color-glass condensate}
\label{sec:cgc}

In the high-energy limit, the eikonal interaction with the target can be described in terms of Wilson lines. This corresponds to particles scattering off a classical color field, which can be written as~\cite{Weigert:2005us,Gelis:2010nm,Kovchegov:2012mbw}
\update{
\begin{equation}
\label{eq:O_average}
    \langle \mathcal{O} \rangle = \int D \rho \mathcal{W}[\rho] \mathcal{O}[\rho]
\end{equation}
}where $\mathcal{O}$ is some combination of Wilson lines, and we average over different color-field configurations of the target with the weight $\mathcal{W}[\rho]$. The weight function can be understood as the probability of finding a given color configuration in the target\update{, and it is normalized such that
\begin{equation}
\label{eq:W_normalization}
    \int D \rho \mathcal{W}[\rho] = 1.
\end{equation}
}Generally, the weight is an unknown nonperturbative function that has to be modeled.
This is usually done with the MV model~\cite{McLerran:1993ka,McLerran:1993ni,McLerran:1994vd} which corresponds to the weight
\begin{equation}
\label{eq:W_MV}
    \mathcal{W}[\rho] = \exp(-\int\dd[3]{z} \frac{ \rhosq(z)}{2\mu^2(z)})
\end{equation}
where the 3-coordinates are given by $z = (z^+,\zt)$ in the coordinate system with the target having a large minus-momentum $P^-$.
Here $z^+$ and $\zt$ are the plus and transverse components of the coordinate, respectively, and similar notation for the coordinates will be used throughout this work.

The MV model can be motivated by the following two main assumptions:
\begin{enumerate}
    \item \label{it:independent} 
    For a heavy nucleus, the color-charge densities of the individual nucleons are largely uncorrelated.
    The nucleons can also be treated as point-like compared to the nucleus, leading to independent probability distributions for the color charge at each point.
    This allows us to write the weight function as:
    \begin{equation}
    \label{eq:weight_independent}
        \mathcal{W}[\rho] \approx \prod_z \mathcal{W}_z[\rho(z)].
    \end{equation}
    where $\mathcal{W}_z$ only depends on the color charge density at the point $z$.
    \item \label{it:random_walk} 
    \update{We have a large number $n$ of color sources over the longitudinal direction.
    This assumption is valid, for example, for a large nucleus where the individual nucleons function as the color sources.} The color charges of these sources are roughly independent of each other, and thus we can model the total color field by a probability density that corresponds to a random walk in the color space~\cite{Jeon:2004rk}, i.e. we can write the probability density as
    \begin{equation}
        p(\vec \rho | n) =  \int \qty[\prod_{i=1}^n 
        \frac{\dd{\Omega_i}}{\vol{S^{D-1}}}
        ] 
        \delta^{(D)}\qty( \vec \rho - \sum_{i=1}^n \vec \rho_i ).
    \end{equation}
    Here $\vec \rho_i$ is the color-field vector of the source $i$,
    $\Omega_i$ denotes the solid angle of the vector $\vec \rho_i$,
    and we have assumed the sources to be identical particles such that for each source the strength $\vec {\rho_i}^2 \equiv \rho_0^2$ is the same constant.
    Thus, we only need to integrate over the angles $\Omega_i$ in the color space with the dimension $D = \nc^2-1$.
    The factor $\vol{S^{D-1}}$ corresponds to the volume of the $(D-1)$-sphere and sets the correct normalization.
    This random walk can be understood as a Wiener process in the color space, and in the limit of a large number of particles we can approximate the probability distribution as a Gaussian:
    \begin{equation}
    \label{eq:rho_fixed_n}
        p(\vec \rho | n) \approx \frac{1}{\qty[2\pi n \rho_0^2 ]^{D/2}} \exp( - \frac{{D \vec \rho}^{\, 2}}{2 n \rho_0^2} ).
    \end{equation}
    This probability distribution can then be identified with the weight $\mathcal{W}$.
\end{enumerate} 
Point~\ref{it:random_walk} shows that for a fixed number of color sources $n$ the Gaussian approximation is valid.
However, for physical nuclei the number of color sources is a fluctuating quantity, and thus we should consider Eq.~\eqref{eq:rho_fixed_n} to actually correspond to the conditional probability $p(\vec \rho | n)$ at a given $n$.
The actual probability distribution for the color densities is then given by
\begin{equation}
    p(\vec \rho) = \int_0^\infty p(\vec \rho | n) p(n)
\end{equation}
where $p(n)$ is the probability to find $n$ color sources in the target.
The distribution for the color field strength $\vec \rho$ is highly dependent on the distribution $p(n)$ and can differ from a Gaussian.
For example, if the distribution $p(n)$ has a heavy tail $p(n) \sim 1/n^{1+\nu}$, then also the distribution $p(\vec \rho)$ behaves like
$p(\vec \rho) \sim 1/\abs{\vec \rho}^{D + 2\nu}$ for large $\rhosq$.
For this reason, we wish to consider what happens if we relax the assumption about the Gaussian probability distribution.

Our model for the weight $\mathcal{W}$ is the following.
First, we will follow Point~\ref{it:independent} and assume that the weights at different points are independent; that is, we will write
\begin{equation}
\label{eq:independent_rho}
    \mathcal{W}[\rho] = \prod_z p_{z}\qty(\rhosq_{z})
\end{equation}
where the probability distributions $p_{z}$ at different points $z$ can be different.
The main motivation for keeping this assumption is that without it analytical calculations are extremely difficult, especially if one assumes that the fields at different \update{longitudinal coordinates} $z^+$ are correlated.
Second, we have also assumed that the probability distributions $p_{z}$ depend only on the square of the color density $\rhosq$.
This assumption can be argued from its invariance under gauge transformations.
However, we note that other combinations of the color density, such as $d^{abc} \rho^a \rho^b \rho^c$, also satisfy gauge invariance.
The reason for considering only the square $\rhosq$ is that it allows us to compute Wilson-line correlators for a very general class of probability distributions, as will be demonstrated in Sec.~\ref{sec:wilson-line_correlators}.

With this model, we will now rewrite Eq.~\eqref{eq:independent_rho} by considering the \textit{characteristic functions} $\varphi_{z}$ instead.
These are defined as the Fourier transforms of the probability distributions $p_{z}$, i.e.
\begin{equation}
\label{eq:p_from_phi}
    p_{z}\qty(\rhosq)
    = \int \dd[D]{\vec \sigma} \qty(\frac{\DD[3]{z}}{2\pi })^D \exp[-i \DD[3]{z} \vec \sigma \vdot \vec \rho]  \varphi_{z}\qty(\sigmasq ).
\end{equation}
Here  the characteristic function $\varphi_{z}$ depends only on the square $\sigmasq$
which follows from the corresponding assumption that the probability distribution $p_z$ also only depends on $\rhosq$.
Additionally, we will rewrite
\begin{equation}
\label{eq:phi}
    \varphi_{z}\qty(\sigmasq) = \exp[ - \DD[3]{z} \w_z\qty(\sigmasq) ]
\end{equation}
which will turn out to be a more useful form later.
Here $\w_z$ is a function that depends on the coordinate $z$ and satisfies $\w_z(0)=0$ for all $z$ to ensure that the probability distributions $p_z$ are properly normalized.
We note that we have included the lattice volume $\DD[3]{z}$ in anticipation of the actual calculations where we will need to expand in terms of the lattice spacing in the plus-direction $\DD{z^+}$.
Note also that in principle the functions $\w_z$ could also have a dependence on the lattice volume.
However, it will turn out that any powers \textit{larger} than $\DD[3]{z}$ are insignificant in the computation, whereas powers smaller than $\DD[3]{z}$ in the exponential will lead to an incorrect continuum limit.
This is our motivation for the inclusion of the lattice spacing $\DD[3]{z}$ in this form into Eqs.~\eqref{eq:p_from_phi} and \eqref{eq:phi}.
Additionally, in the continuum limit $\DD[3]{z} \to 0$ our weight function will have the appropriate form
\begin{equation}
\label{eq:W_general}
    \mathcal{W}[\rho] = \int \dcal{\sigma} 
    \exp\qty[ -i \int \dd[3]{z} \vec \sigma_z \vdot  \vec \rho_z -\int \dd[3]{z} \w_z\qty(\sigmasq_z)   ].
\end{equation}
Here the factors of $\DD[3]{z}/(2\pi)$ have been absorbed into the definition of the functional integration measure $\dcal{\sigma} = \prod_z \prod_a [(\dd{\sigma_z^a} \times \DD[3]{z}/(2\pi) ] $
\update{
to satisfy the normalization condition Eq.~\eqref{eq:W_normalization}.}
Equation~\eqref{eq:W_general} defines our model for the weight function used for computations in Sec.~\ref{sec:wilson-line_correlators}, and in Sec.~\ref{sec:stable} we will consider a specific form for the function $w_z$.

\section{Computing Wilson-line correlators}
\label{sec:wilson-line_correlators}

\subsection{Dipole amplitude}
Perhaps the most central quantity that appears in calculations in the high-energy limit is the dipole amplitude, defined as
\begin{equation}
\label{eq:dipole_amplitude}
    N(\xt, \yt) = 1 -  \expval{S^{(2)}(\xt,\yt)},
\end{equation}
where we have defined the dipole--target scattering matrix
\begin{equation}
    S^{(2)}(\xt,\yt) \equiv \frac{1}{\nc} \Tr[V(\xt) V^\dag(\yt)],
\end{equation}
and
\begin{equation}
\label{eq:wilson_line}
    V(\xt) = \mathcal{P} \exp( -i \gs \int_{-\infty}^\infty \dd{x^+} A^{a-}(\xt, x^+) t^a )
\end{equation}
is a Wilson line in the fundamental representation. 
The notation $\mathcal{P}$ in the Wilson line~\eqref{eq:wilson_line} means that it is a path-ordered exponential, with $x^+$ increasing from right to left.
The dipole amplitude is the simplest correlator that one encounters in physical computations: 
the correlator of a single Wilson line $\expval{ \Tr[V(\xt)]}$, while being simpler, does not have any practical use in measurable quantities.
The gluon field $A^{-}$ is related to the classical color field $\rho$ by
\begin{equation}
\label{eq:rho_from_A}
    \qty(-\boldsymbol{\nabla}^2 + m^2) A^{a-} = \rho^a
\end{equation}
where we also include an infrared regulator $m$.
The solution to this equation can be written as
\begin{equation}
   A^{a-}(\xt,x^+) =\frac{1}{2\pi} \int \dd[2]{\ut} K_0( m \abs{\xt-\ut} )  \rho^a(\ut,x^+),
\end{equation}
assuming $A^{a-}(\xt) \to 0$ for large transverse coordinates $\xt$.

We will now use our assumption~\eqref{eq:independent_rho} that the probability densities for color fields at different \update{longitudinal coordinates} $z^+$ are not correlated.
As the Wilson lines are path-ordered exponentials in the \update{longitudinal coordinate}, this means that we can calculate the target averages
\begin{equation}
\label{eq:xplus_wilson_line}
  S_{z^+}^{ijkl} =   \left \langle \exp( -i \gs \DD{z^+} A^{a-}_{\xt} t^a )_{ij} \exp( i \gs \DD{z^+} A^{b-}_{\yt} t^b )_{kl}  \right \rangle_{z^+}
\end{equation}
separately for each $z^+$.
Here the subscripts $ijkl$ denote the color indices of the matrix exponentials.
The standard derivation of the dipole amplitude in the MV model then follows from expanding the exponentials in Eq.~\eqref{eq:xplus_wilson_line} in terms of $\DD{z^+}$ and keeping only the leading terms.
This approach allows us to get the dipole amplitude by multiplying the ($\DD{z^+}$-expanded) correlators at different \update{longitudinal coordinates} $z^+$, with the result being exact due to the higher-order terms in $\DD{z^+}$ being suppressed.
However, for our general model~\eqref{eq:W_general} this expansion may break down as the averages
$\expval{ A^{a-}_\xt A^{b-}_\yt }$
are not guaranteed to be finite.
At the same time, the original expectation value~\eqref{eq:xplus_wilson_line} is finite for \textit{any} collection of probability distributions $p_z$, which follows from the matrix exponentials being bounded as unitary matrices.
For this reason, we will follow a different approach in this work that works more generally:
instead of expanding the exponential matrices in Eq.~\eqref{eq:xplus_wilson_line} in terms of $\DD{z^+}$, we will do this expansion for the characteristic function $\varphi_z$~\eqref{eq:phi}.
This way, each term in the expansion is well-defined.

Expanding the characteristic function $\varphi_z$ to the leading nontrivial order, we find
\begin{equation}
\label{eq:W_expansion}
    \mathcal{W} \qty[\rho]
    =  \int \dcal{\sigma} e^{-i \int \dd[3]{z} \vec \rho_z \vdot \vec \sigma_z  } 
  \prod_{z^+}  \qty{ 1 -  \DD{z^+} \int \dd[2]{\zt} \w_z\qty( \sigmasq_z) + \order{\qty[\DD{z^+}]^2}  }.
\end{equation}
This leads to a significant simplification: note that for the first term in the expansion in Eq.~\eqref{eq:W_expansion} we can trivially integrate over $\sigma_z$ for all $\zt$, giving us delta functions that set $\rho_{z^+} \to 0$.
The same can be done for the second term, \textit{except} for a single value of $\zt$ under the transverse integral.
Integrating over the delta functions of $\rho$,
we get
\begin{equation}
    \label{eq:xplus_expanded}
    \begin{split}
         S_{z^+}^{ijkl}
    =&
    \delta^{ij} \delta^{kl} 
    -
    \DD{z^+}
    \int \dd[2]{\zt}
    \int \dd[D]{\vec \rho} \dd[D]{\vec \sigma} \qty(\frac{\DD[3]{z}}{2\pi})^D
    e^{- i \DD[3]{z} \vec \rho \vdot \vec \sigma }
    \w_z\qty(\sigmasq)
    U_{\xt-\zt}\qty(\vec\rho)_{ij}
    U^\dag_{\yt-\zt}\qty(\vec\rho)_{kl}
    \\
    &+\order{\qty[\DD{z^+}]^2} 
    \end{split}
\end{equation}
where we have defined the unitary matrix
\begin{equation}
    U_{\xt-\zt}\qty(\vec\rho)
    =\exp( - i \frac{\gs \DD[3]{z}}{2\pi} K_0(m \abs{\xt-\zt}) \rho^a t^a ).
\end{equation}

Let us now focus on computing the term proportional to $\DD{z^+}$.
First, we note that when computing the trace for the lowest (or highest) value of $z^+$ we will contract the indices $il$ (or ${jk}$) together.
The remaining structure then has to be proportional to $\delta^{jk}$ (or $\delta^{il}$), which can be seen by considering the possible combinations of color indices after integration.
This same structure in color indices propagates to all values of $z^+$, meaning that we only need to compute the structure
$S_{z^+} \equiv \frac{1}{\nc} S_{z^+}^{ijji}$.
Second,
we can integrate over the angle of the vector $\vec \sigma$ analytically by using the identity
\begin{equation}
    \int \dd[D]{\vec \sigma} e^{- i \vec \rho \vdot \vec \sigma} f\qty(\sigmasq)
    = \rho \int_0^\infty \dd{\sigma} \qty( \frac{2\pi \sigma}{ \rho  })^{\frac{D}{2}} J_{D/2-1}\qty(\sigma \rho)
    f(\sigma^2)
\end{equation}
where $\sigma = \sqrt{ \sigmasq }$ and $\rho = \sqrt{\rhosq}$, and $J$ is the Bessel function of the first kind.
Third,
we note that we can now combine the exponentials
\begin{equation}
    U_{\xt-\zt}\qty(\vec\rho)
    U^\dag_{\yt-\zt}\qty(\vec\rho)
     =
    \exp( - i \frac{\gs \DD[3]{z}}{2\pi} K_{\xt \yt \zt} \rho^a t^a)
\end{equation}
where we have denoted 
$K_{\xt \yt \zt} = K_0(m \abs{\xt -\zt}) - K_0(m \abs{\yt -\zt})$.
Fourth, we rescale 
\begin{align}
  \frac{\gs\DD[3]{z}}{2\pi} K_{\xt \yt \zt} \rho &\to  \rho,
  &
  \sigma &\to \frac{\gs}{2\pi} K_{\xt \yt \zt} \sigma,  
\end{align}
in the integration
and write the vector $\vec \rho$ in terms of the norm $\rho$ and the angular part $\Omega_\rho$.
In the end, we get
\begin{equation}
\label{eq:S_general}
\begin{split}
    S_{z^+}
    =& 1 - \DD{z^+} \int \dd[2]{\zt} \int_0^\infty \dd{\sigma} \dd{\rho} \int_{S^{D-1}} \dd{\Omega_\rho} \qty(  \frac{\sigma \rho}{2\pi} )^{\frac{D}{2}} J_{D/2-1}(\sigma \rho) 
    \w_z\qty(\qty[\frac{\gs}{2\pi} K_{\xt \yt \zt}\sigma]^2) \\
    &\times
    \frac{1}{\nc}\Tr\qty\Big[ \exp( -i  \rho \hat \rho^a t^a   ) ]
    +\order{\qty[\DD{z^+}]^2}.
\end{split}
\end{equation}
Here $\hat \rho^a = \rho^a/ \rho$ denotes the unit vector in the direction of $\vec \rho$, and it depends only on the solid angle $\Omega_\rho$.

Equation~\eqref{eq:S_general} is as far as we can go \update{using the weight in Eq.~\eqref{eq:W_general}} without specifying the number of colors $\nc$, along with the function $\w_z$ determining the exact model for the color density.
Once $\nc$ is chosen, one can then proceed to integrate over the angle $\Omega_\rho$.
This can be done by noting that the angular dependence is contained in $\hat \rho$, and using representation theory this integral can be evaluated analytically for any representation and number of colors.
The general procedure for evaluating the integral is explained in Appendix~\ref{app:moment-generating_function}.
This allows us to compute the function
\begin{equation}
\label{eq:xi}
    \xi_R(\sigma) = \int_0^\infty \dd{\rho} \int_{S^{D-1}} \dd{\Omega_\rho} \qty(\frac{\sigma \rho}{2\pi})^{\frac{D}{2}}
    J_{D/2-1}\qty(\sigma \rho) \times \frac{1}{D_R} \Tr\qty[ \exp( -i t_R^a \hat \rho^a \rho ) ]
\end{equation}
which we have defined for a general representation $R$, with color matrices $t_R^a$ and dimension $D_R$.
We note that, because of the factor $\rho^{\frac{D}{2}}$, this integral generally converges only in the distributional sense.
We will also define the function
\begin{equation}
\label{eq:F_z}
    F^R_z(\xt, \yt) =  \int_0^\infty \dd{\sigma} \xi_R(\sigma) \,  \w_z\qty(\qty[\frac{\gs}{2\pi} K_{\xt \yt \zt}\sigma]^2) 
\end{equation}
which depends on the model $\w_z$.
Equation~\eqref{eq:S_general} can then be rewritten as
\begin{equation}
\label{eq:S_expanded}
    S_{z^+} = 1 - \DD{z^+} \int \dd[2]{\zt} F^R_{z}(\xt,\yt) +\order{\qty[\DD{z^+}]^2}.
\end{equation}
Exponentiating this result gives us the dipole amplitude
\begin{equation}
\label{eq:N_solved}
 N(\xt, \yt) = 1 - \prod_{z^+ }S_{z^+}   
 = 1 - \exp( - \int \dd[3]{z} F^R_{z}(\xt, \yt) )
\end{equation}
where we have also taken the continuum limit $\DD{z^+} \to 0$.
Note that this exponentiation relies on the correct dependence on $\DD{z^+}$ in Eq.~\eqref{eq:S_expanded}, as any other power than one would not lead to a well-defined $z^+$-integral in Eq.~\eqref{eq:N_solved}.
Terms $\order{\qty[\DD{z^+}]^2}$ in $S_{z^+}$ can be dropped in the exponentiation, and the result~\eqref{eq:N_solved}
is exact in the continuum limit.

The computation outlined in this section is quite general: the dipole amplitude in \textit{any} representation $R$ can be computed using this procedure, and in fact Eq.~\eqref{eq:N_solved} is valid for any $R$.
All of the information about the representation is included in the function $\xi_R$ defined in Eq.~\eqref{eq:xi}.

\subsection{Higher-order correlators}
\label{sec:higher-order}

The weight function~\eqref{eq:W_general} can also be used to compute higher-order correlators of Wilson lines.
However, the computation becomes much more involved as 
the color indices for each individual $z^+$ cannot be directly summed over anymore, meaning that one has to compute tensor exponentials with non-trivial contractions~\cite{Blaizot:2004wv,Dominguez:2008aa,Dominguez:2011wm,Kovner:2001vi}.
This is also true for the Gaussian model, and for this reason higher-order correlators are more easily computed by forming and solving a differential equation instead~\cite{Lappi:2017skr,Isaksen:2020npj}.
The same idea can be applied to the more general weight~\eqref{eq:W_general}.

Let us formulate the general idea here.
We will look at the correlator
\begin{equation}
    S_\alpha(L^+) = \expval{ \qty[\otimes_i V^{R_i}(\xt_i)]_\alpha }_{(-\infty,L^+)}
\end{equation}
where $V^R$ is a Wilson line in representation $R$, and $\alpha$ denotes the general color indices of this collection of Wilson lines.
The notation $ \expval{\ldots}_{(-\infty,L^+)}$ corresponds to an expectation value that is computed using the weight
\begin{equation}
    \mathcal{W}_{(-\infty,L^+)}\qty[\rho]
    = 
    \int \dcal{\sigma} 
    \exp\qty[ -i \int_{-\infty}^{\infty} \dd{z^+} \int \dd[2]{\zt} \vec \sigma_z \vdot \vec \rho_z -\int_{-\infty}^{L^+} \dd{z^+}\int \dd[2]{\zt} \w_z\qty(\sigmasq_z)   ]
\end{equation}
such that we have set an upper limit $L^+$ for the integral with the function $\w_z$.
This integral sets the density $\rho_z$ to zero at \update{longitudinal coordinates} $z^+ > L^+$.
We can then compute the derivative
\begin{equation}
\label{eq:S_der}
\begin{split}
    \dv{S_\alpha(L^+)}{L^+}
    =&
   -\int  \dcal{\rho} \dcal{\sigma}
      \exp\qty[ -i \int_{-\infty}^{\infty} \dd{z^+} \int \dd[2]{\zt} \vec \sigma_z \vdot \vec \rho_z -\int_{-\infty}^{L^+} \dd{z^+}\int \dd[2]{\zt} \w_z\qty(\sigmasq_z)   ]
  \\
   &\times     \qty[\otimes_i V^{R_i}(\xt_i)]_\alpha
    \times \int \dd[2]{\zt} \w_{(L^+, \zt)}\qty(\sigmasq_{(L^+, \zt)})  
    .
\end{split}
\end{equation}
This modifies the functional integral only for the \update{longitudinal coordinate} $L^+$, and the expectation values for the other \update{longitudinal coordinates} give us $S_\beta(L^+)$ with some color indices $\beta$.
For the \update{coordinate} $L^+$ we only need to keep the leading term in the expansion in $\DD{z^+}$, and we can simplify Eq.~\eqref{eq:S_der} to
\begin{equation}
\begin{split}
    \dv{S_\alpha(L^+)}{L^+}
     =&- S_{\beta}(L^+)
     \times
     \int \dd[2]{\zt}
     \int \dd[D]{\vec\rho} \dd[D]{\vec\sigma} \qty( \frac{\DD[3]{z}}{2\pi} )^D 
     e^{- i \DD[3]{z} \vec \rho \vdot \vec \sigma }
     \times \w_{(L^+, \zt)}\qty(\sigmasq)    \\
     & \times  \qty[  \otimes_i \exp( - i \frac{\gs \DD[3]{z}}{2\pi} K_0(m \abs{\zt -\xt_i}) \rho^a t_{R_i}^a )]_{\alpha \beta} \\
     =&-
     S_{\beta}(L^+)
     \times 
     \int \dd[2]{\zt}
     \int_0^\infty  \dd{\sigma} \dd{\rho} \int_{S^{D-1}} \dd{\Omega_\rho}
     \qty(\frac{\sigma \rho}{2\pi})^D
     J_{D/2-1}(\sigma \rho)
     e^{- i \vec \rho \vdot \vec \sigma } \w_{(L^+, \zt)}\qty(\sigma^2)    \\
     & \times  \qty[  \otimes_i \exp( - i \frac{\gs}{2\pi} K_0(m \abs{\zt -\xt_i}) \rho \hat \rho^a t_{R_i}^a )]_{\alpha \beta}
\end{split}
\end{equation}
where we sum over the color indices $\beta$.
This has the form of a matrix differential equation
\begin{equation}
    \dv{S_\alpha(L^+)}{L^+}
    = M_{\alpha \beta}(L^+) S_\beta(L^+)
\end{equation}
which can be solved.
The initial condition at $L^+ = -\infty$ corresponds to setting $V^R_{ij} \to \delta^{ij}$ for all Wilson lines, and the solution at $L^+ = \infty$ gives us the desired Wilson-line correlator.

In practice, the difficult part is computing the matrix $M_{\alpha \beta}$.
If we only consider Wilson lines in the fundamental representation without additional color matrices $t^a$, we can decompose the matrix $M_{\alpha \beta}$ to Kronecker deltas involving different combinations of the multi-indices $\alpha$ and $\beta$ along with traces of the exponential matrices.
This allows us to use the computational framework in Appendix~\ref{app:moment-generating_function}.
For other representations, the color structure of the matrix $M_{\alpha \beta}$ can be more general than simple Kronecker deltas, making analytical computations more difficult.
However, if we only consider physical quantities with only fundamental and adjoint Wilson lines, the color indices must be contracted with each other, allowing us to reduce the Wilson-line correlator into traces of fundamental Wilson lines only.
Thus, for physical quantities the relevant Wilson-line correlators can be evaluated analytically when using the weight function of Eq.~\eqref{eq:W_general}.

\subsection{Crosscheck: adjoint dipole amplitude and the 4-point correlator}

\label{sec:stable_4}

As a check of the validity of the current framework for computing higher-order correlators, let us consider the adjoint dipole amplitude.
This is given by Eq.~\eqref{eq:N_solved} with the function $F^R_z$ now being different from the fundamental representation.
The crucial thing here is that the adjoint dipole amplitude can also be written as a correlator of four fundamental Wilson lines, allowing us to compare the more direct result~\eqref{eq:N_solved} with one computed using the method outlined in Sec.~\ref{sec:higher-order}.
To see this, note that we can write the adjoint Wilson line as
\begin{equation}
    \label{eq:adjoint_Wilson}
    U^{ab}(\xt) = 2\Tr[ t^a V(\xt)t^b V^\dag(\xt) ]
\end{equation}
where $V$ is a Wilson line in the fundamental representation.
The adjoint dipole amplitude is given by
\begin{equation}
    \label{eq:adjoint_dipole_amplitude}
    N_A(\xt,\yt)
    = 1 -  \expval{S_A^{(2)}(\xt,\yt)}
    = 1 - \frac{1}{\nc^2-1} \expval{\Tr[ U(\xt) U^\dag(\yt) ] },
\end{equation}
and  we can rewrite the adjoint scattering matrix as
\begin{equation}
\begin{split}
   S_A^{(2)}(\xt,\yt) 
   =& \frac{1}{\nc^2-1} \qty{\Tr[V(\xt)V^\dag(\yt)]\Tr[V(\yt)V^\dag(\xt)]  -1}\\
   =& \frac{1}{\nc^2 - 1}
   \qty{ \nc^2 S^{(2)}(\xt,\yt) S^{(2)}(\yt,\xt) - 1 }    
\end{split}
\end{equation}
by using the Fierz identity
\begin{equation}
    \label{eq:fierz}
    t^a_{ij} t^a_{kl} = \frac{1}{2} \qty[ \delta^{il} \delta^{jk} - \frac{1}{\nc} \delta^{ij} \delta^{kl}].
\end{equation} 
This implies that we must have
\begin{equation}
\label{eq:4point}
     \expval{ S^{(2)}(\xt,\yt) S^{(2)}(\yt,\xt)}
    =\frac{1}{\nc^2} + \frac{\nc^2-1}{\nc^2} \exp\qty{
    - 
    \int \dd[3]{z} F^A_z(\xt,\yt)
    }
\end{equation}
where we have denoted by $ F^A$ the function $F^R$ in the adjoint representation.
We now want to check Eq.~\eqref{eq:4point} by directly considering the double-dipole correlator $ \expval{S^{(2)} S^{(2)}}$.

As outlined in Sec.~\ref{sec:higher-order}, we can build a differential equation to evaluate the 4-point correlator.
We end up with the following differential equation:
\begin{equation}
\label{eq:4DE}
\begin{split}
\frac{\dd}{\dd{L^+}}
  &\expval{ S^{(2)}(\xt,\yt) S^{(2)}(\yt,\xt) }_{(-\infty,L^+)}\\
=&-
\frac{1}{\nc^2}  \expval{   \qty[V(\xt)V^\dag(\yt)]_{ij} \qty[V(\yt)V^\dag(\xt)]_{kl} }_{(-\infty,L^+)} 
 \\ 
 & \times \int \dd[2]{\zt}
 \int_0^\infty \dd{\sigma}\dd{\rho}
 \int_{S^{D-1}} \dd{\Omega_\rho} \qty(\frac{\sigma \rho}{2\pi})^D J_{D/2-1}(\sigma \rho) 
 e^{- i \vec \rho \vdot \vec \sigma}
 w_{(L^+,\zt)}\qty(\sigma^2) \\
&\times  \qty[\exp( -i \frac{\gs}{2\pi} K_{\xt \yt \zt} \rho \hat \rho^a t^a )]_{ji}
 \qty[\exp( +i \frac{\gs}{2\pi} K_{\xt \yt \zt} \rho \hat \rho^a t^a )]_{lk}  .
\end{split}
\end{equation}
Denoting
$\mathcal{U} = \exp( -i \frac{\gs}{2\pi} K_{\xt \yt \zt} \rho \hat \rho^a t^a )$,
the last line has the structure $\mathcal{U}_{ji} \mathcal{U}^{-1}_{lk}$.
After the integration over $\Omega_\rho$, this quantity cannot explicitly depend on the color matrices $t^a$ but instead can only have the following color structure:
\begin{equation}
    \mathcal{U}_{ji} \mathcal{U}^{-1}_{lk} = c_S \delta^{ji} \delta^{lk} + c_A \delta^{jk} \delta^{li}
\end{equation}
where $c_S$ and $c_A$ are coefficients that can be written in terms of traces of $\mathcal{U}$.
Explicitly, we find
\begin{equation}
    \begin{split}
        c_S &= \frac{1}{\nc^2 - 1} \qty{ \Tr[\mathcal{U}] \Tr[\smash{\mathcal{U}^{-1}}] - 1   },\\
        c_A &= \frac{1}{\nc^2 - 1} \qty{ -\frac{1}{\nc} \Tr[\mathcal{U}] \Tr[\smash{\mathcal{U}^{-1}}] + \nc  }.
    \end{split}
\end{equation}
Equation~\eqref{eq:4DE} can then be written as
\begin{equation}
\label{eq:4DE2}
\begin{split}
\frac{\dd}{\dd{L^+}}
  &\expval{S^{(2)}(\xt,\yt) S^{(2)}(\yt,\xt)  }_{(-\infty,L^+)}\\
=&-
 \int \dd[2]{\zt}
 \int_0^\infty \dd{\sigma}\dd{\rho}
 \int_{S^{D-1}} \dd{\Omega_\rho} \qty(\frac{\sigma \rho}{2\pi})^D J_{D/2-1}(\sigma \rho) 
 e^{- i \vec \rho \vdot \vec \sigma}
 w_{(L^+,\zt)}\qty(\sigma^2) \\
&\times  
\qty{
c_S \expval{S^{(2)}(\xt,\yt) S^{(2)}(\yt,\xt)  }_{(-\infty,L^+)}
+ \frac{c_A}{\nc}
} \\
=&-
 \int \dd[2]{\zt}
 \int_0^\infty \dd{\sigma}\dd{\rho}
 \int_{S^{D-1}} \dd{\Omega_\rho} \qty(\frac{\sigma \rho}{2\pi})^D J_{D/2-1}(\sigma \rho) 
 e^{- i \vec \rho \vdot \vec \sigma}
 w_{(L^+,\zt)}\qty(\sigma^2) \\
&\times  
\frac{1}{\nc^2 - 1 } 
 \Tr[\mathcal{U}] \Tr[\smash{\mathcal{U}^{-1}}]
\qty{
 \expval{S^{(2)}(\xt,\yt) S^{(2)}(\yt,\xt)  }_{(-\infty,L^+)}
-\frac{1}{\nc^2}
} 
\end{split}
\end{equation}
where in the last step we have used the fact that for the constant terms in $c_S$ and $c_A$ we can directly integrate over $\vec \rho$, giving us a delta function $\delta^{(D)}( \vec \sigma )$.
This contribution then vanishes since $w(\sigma^2) = 0$.

As outlined in Appendix~\ref{app:moment-generating_function}, 
we have
\begin{equation}
\Tr[\mathcal{U}] \Tr[\smash{\mathcal{U}^{-1}}]
 =
 \chi_F(\vec \lambda)
 \chi_{\overline F}(\vec \lambda)
 = 1 + \chi_A(\vec \lambda)
\end{equation}
 and thus this agrees with the expression for the adjoint representation, $\chi_A(\vec \lambda)$, up to a constant term.
However, as explained above, the contribution of the constant vanishes, and we can use the result for the adjoint representation.
The differential equation reduces to 
\begin{equation}
\label{eq:4DE3}
\begin{split}
\frac{\dd}{\dd{L^+}}
  &\expval{S^{(2)}(\xt,\yt) S^{(2)}(\yt,\xt)   }_{(-\infty,L^+)}\\
=&-
\qty{
 \expval{S^{(2)}(\xt,\yt) S^{(2)}(\yt,\xt)   }_{(-\infty,L^+)}
-\frac{1}{\nc^2}
}
\times
\int \dd[2]{\zt} F^A_z(\xt,\yt)
 ,
\end{split}
\end{equation}
and taking into account the initial condition $\expval{S^{(2)}(\xt,\yt) S^{(2)}(\yt,\xt) }_{(-\infty,-\infty)} =1$ this is solved by 
\begin{equation}
\label{eq:4point_L}
     \expval{ S^{(2)}(\xt,\yt) S^{(2)}(\yt,\xt)}_{(-\infty,L^+)}
    = \frac{1}{\nc^2} + \frac{\nc^2-1}{\nc^2} \exp\qty{
    - 
    \int_{-\infty}^{L^+} \dd{z^+} \int \dd[2]{\zt} F^A_z(\xt,\yt)
    },
\end{equation}
in agreement with Eq.~\eqref{eq:4point} when  $L^+ = +\infty$.

This example of solving a higher-order correlator using a differential equation is one of the simpler cases.
Had we considered a correlator of two dipoles with different coordinates, we would have ended up with terms that are also proportional to the quadrupole operator
\update{
\begin{equation}
S^{(4)}(\xt, \yt, \xt', \yt') = \frac{1}{\nc} \Tr[V(\xt)V^\dag(\yt)V(\xt')V^\dag(\yt')] .   
\end{equation}
}
In general, this leads to the following coupled differential equation:
\begin{equation}
    \frac{\dd}{\dd{L^+}}
    \Biggl\langle
    \mqty(
S^{(2)}(\xt, \yt) S^{(2)}(\xt', \yt') \\ 
S^{(2)}(\xt, \yt') S^{(2)}(\xt', \yt) \\ 
S^{(4)}(\xt, \yt, \xt', \yt') \\ 
S^{(4)}(\xt, \yt', \xt', \yt) 
)
\Biggr\rangle_{(-\infty,L^+)}
= M(L^+) \,
    \Biggl\langle
    \mqty(
S^{(2)}(\xt, \yt) S^{(2)}(\xt', \yt') \\ 
S^{(2)}(\xt, \yt') S^{(2)}(\xt', \yt) \\ 
S^{(4)}(\xt, \yt, \xt', \yt') \\ 
S^{(4)}(\xt, \yt', \xt', \yt) 
)
\Biggr\rangle_{(-\infty,L^+)}
\end{equation}
where 
the matrix $M(L^+)$ can depend on $L^+$ and the coordinates.
This can be formally solved with
\begin{equation} \Biggl\langle
    \mqty(
S^{(2)}(\xt, \yt) S^{(2)}(\xt', \yt') \\ 
S^{(2)}(\xt, \yt') S^{(2)}(\xt', \yt) \\ 
S^{(4)}(\xt, \yt, \xt', \yt') \\ 
S^{(4)}(\xt, \yt', \xt', \yt) 
)
\Biggr\rangle_{(-\infty,L^+)}
= \mathcal{P} \exp( \int_{-\infty}^{L^+} \dd{z^+}M(z^+))
    \mqty(
1 \\ 1 \\ 1 \\ 1 
)
\end{equation}
where $\mathcal{P}$ denotes a time-ordered matrix exponential from right to left.
This expression is analogous to the ones used in the Gaussian approximation to compute higher-order correlators.
Note that the time-ordering introduces a non-trivial dependence on the functional form of $w_z$ in terms of the \update{longitudinal coordinate} $z^+$:
while for the dipole amplitude~\eqref{eq:N_solved} only the integrated form $\int \dd{z^+} w_z$ appears, this is not necessarily true for higher-order correlators.
The fact that the matrices $M(z^+)$ at different \update{coordinates} $z^+$ do not generally commute means that the higher-order correlators can have a more complicated dependence on $w_z$, and the shape of the function $w_z$ in terms $z^+$ could potentially matter.
This observation is true even for the Gaussian case.
However, although we have not checked this explicitly, this dependence is expected to be small as it only appears for higher-order correlators through the difference in the transverse shape of $w_z$ for different \update{longitudinal coordinates} $z^+$.

\subsection{From nucleons to nuclei}

Even though the assumptions made about the high gluon density are more valid for heavy nuclei than for protons, a sizable amount of the current data is for proton targets.
This is especially true for deep inelastic scattering which provides the most straightforward way of constraining the dipole amplitude and thus the color density of the target.
For this reason,
it is more practical to fit the model for proton targets and then try to describe the nucleus as a collection of nucleons.
We wish to consider how this can be done at the level of the color density.

Let us begin by assuming that for nucleons we have the color density of Eq.~\eqref{eq:W_general}, determined by the function $\w_z$, 
and the location of a nucleon inside the nucleus is given by a probability distribution $p_n(z)$.
We can then write the weight $\mathcal{W}_A$ for a nucleus with a mass number $A$ as
\begin{equation}
\label{eq:W_nucleus}
\mathcal{W}_A[\rho]
=   \prod_{i=1}^A \qty\Bigg[\int \dcal{\rho_i} \int \dd[3]{z_i} p_{n}(z_i) \mathcal{W}_{n_i}[\rho_i] ] 
\times \delta^{(D)}\qty[\vec \rho - \sum_{i=1}^A \vec \rho_i ],
\end{equation}
where the nucleons are assumed to be distributed independently inside the nucleus.
Writing the nucleon weights $\mathcal{W}_i$ in terms of their characteristic functions, we can evaluate the functional integrals over $\rho_i$. 
This leads to
\begin{equation}
\label{eq:W_nucleus_FT}
    \mathcal{W}_A[\rho] =
     \prod_{i=1}^A \qty\Bigg[ \int \dd[3]{z_i} p_{n}(z_i)  ]
    \int \dcal{\sigma} 
    \exp\qty[ -i \int \dd[3]{z} \vec \sigma_z \vdot \vec \rho_z -
    \sum_{i=1}^A \int \dd[3]{z} \w_{z-z_i}\qty(\sigmasq_z)   ].
\end{equation}
The nucleonic result is then changed to
\begin{equation}
\label{eq:proton_to_nucleus}
    \exp[ - \int \dd[3]{z} \w_z \qty(\sigmasq_z) ]
    \to
    \prod_{i=1}^A 
    \int \dd[3]{z_i} p_n(z_i)
     \exp[ - \int \dd[3]{z} \w_{z-z_i} \qty(\sigmasq_z) ]
\end{equation}
for a heavy nucleus.
We can then follow the derivation for the dipole amplitude --- or even higher-order correlators --- and note that there is no significant change in the derivation using this weight function.
Crucially, we can again expand the exponential in terms of $\DD{z^+}$ and only need to keep the first term in the expansion.
Thus, we only need to keep
\begin{equation}
     \sum_{i=1}^A
    \int \dd[3]{z_i} p_n(z_i)
     \DD{z^+} \int \dd[2]{\zt} \w_{z-z_i} \qty(\sigmasq_z)
     =
       \DD{z^+} \int \dd[2]{\zt} \w^A_{z} \qty(\sigmasq_z)
\end{equation}
where we have defined the nuclear color density
\begin{equation}
    \w^A_{z} \qty(\sigmasq_z) \equiv
      \sum_{i=1}^A
    \int \dd[3]{z_i} p_n(z_i)
    \w_{z-z_i} \qty(\sigmasq_z) 
    =
     A \int \dd[3]{z_0} p_n(z_0)
    \w_{z-z_0} \qty(\sigmasq_z)
    .
\end{equation}
The nuclear weight function can then be written in the simple form
\begin{equation}
\label{eq:W_nucleus_final}
    \mathcal{W}_A[\rho] =
    \int \dcal{\sigma} 
    \exp\qty[ -i \int \dd[3]{z} \vec \sigma_z \vdot \vec \rho_z -
  \int \dd[3]{z} \w_{z}^A\qty(\sigmasq_z)   ],
\end{equation}
which agrees with Eq.~\eqref{eq:W_nucleus} up to corrections that vanish in the continuum limit.

It is interesting to consider this result in the limit where the nucleons are much smaller than the nucleus.
This means that $\w_z$ is highly peaked compared to $p_n$, with the peak at $z=(0^+, \boldsymbol{0})$ corresponding to the center of the nucleon. 
This allows us to approximate
\begin{equation}
    \label{eq:smooth_nucleus}
    \w_z^A(\sigmasq)
    \approx A p_n(z) 
    \int \dd[3]{z_0} \w_{z_0}(\sigmasq),
\end{equation}
which agrees with the optical Glauber approach for modeling a heavy nucleus~\cite{Kowalski:2003hm}.

\section{Stable color-glass condensate}
\label{sec:stable}

In this Section, we will consider a specific model for the function $\w_z$ determining the color density to illustrate practical computations in this general approach.
For our model, we choose
\begin{equation}
\label{eq:stable}
    \w_z \qty(\sigmasq) = \mu_z^2 \sigma^{\alpha},
\end{equation}
with $0 < \alpha \leq 2$ and $\mu_z$ being a general function of $z$.
This model is motivated by the following observations:
\begin{enumerate}
    \item For the function $\w_z$ to define a properly normalized probability distribution $p_z(\rho)$, we must have $\w_z(0)=0$.
    Thus, $\alpha > 0$.
    \item For the function $\w_z$ to define a positive definite probability distribution $p_z(\rho_z)$, the small-$\sigma$ behavior $\w_z(\sigmasq)$ must be at most $\order{\sigma^2}$.
    Otherwise, for $\alpha > 2$, we would have
    \begin{equation}
        \int \dd[D]{\vec \rho} \rho^2 p_z(\rho)
        \propto - \int \dd[D]{\vec \sigma}
        \delta^{(D)}(\vec \sigma)
      \times   \nabla^2_\sigma \exp[- \DD[3]{z}  \mu_z^2 \sigma^{\alpha}]
         =0
    \end{equation}
    which is impossible for a proper probability distribution.
    Thus, we must have $\alpha \leq 2$.
    Note that the case $\alpha=2$ corresponds to the standard Gaussian model.
    \item \label{point:power_law} Assume that the probability distribution $p_z$ depends only on a single dimensionful scaling parameter $\mu^2$ whose dimension we take to be $\qty[\si{GeV^{3}}]$, corresponding to a density.
    By dimensional analysis, we must then have
    \begin{equation}
        p_z\qty(\rhosq)
        = 
          \qty[\mu^2 \DD[3]{z} ]^{D /\alpha}
        \qty[\DD[3]{z}]^{D} 
        \times
        \hat p\qty(  \qty[\mu^2 \DD[3]{z} ]^{1/\alpha} \DD[3]{z}  \rho )
    \end{equation}
    for some constant $\alpha$,
    where $\hat p$ only depends on dimensionless quantities and the dependence on $\mu^2$ or $\DD[3]{z}$ only appears through its argument.
    The corresponding characteristic function behaves as
    \begin{equation}
    \label{eq:phi_dimensionless}
        \varphi\qty(\sigmasq)
        = \hat \varphi\qty(  \qty[\mu^2 \DD[3]{z} ]^{1/\alpha} \sigma )
        = \exp\qty[
        - \sum_{i=0} C_i \qty( \qty[\mu^2 \DD[3]{z} ]^{1/\alpha} \sigma )^{n_i}
        ]
    \end{equation}
    where $\hat \varphi$ again depends only on dimensionless quantities, and we have assumed that we can expand the term in the exponential as a power series with powers $n_i$ and coefficients $C_i$.
    We note that to have a proper continuum limit $\DD[3]{z} \to 0$ in Sec.~\ref{sec:wilson-line_correlators}, the smallest power must be $n_i =\alpha$, and any powers higher than this are irrelevant in the continuum limit.
    Equation~\eqref{eq:phi_dimensionless} then reduces to 
    \begin{equation}
           \varphi\qty(\sigmasq)
        =\exp[ - C_0 \mu^2 \DD[3]{z} \sigma^\alpha  ]
    \end{equation}
    which gives us the model in Eq.~\eqref{eq:stable} after absorbing the value of $C_0$ to $\mu^2$.
\end{enumerate}
The model in Eq.~\eqref{eq:stable} also has an interpretation in terms of a stable distribution in higher dimensions.
These distributions satisfy a generalized version of the central limit theorem, stating that the sum of identically distributed random variables tends to a limiting distribution based on the behavior of the tail of the probability distribution.
In this sense, Eq.~\eqref{eq:stable} is a natural extension of the standard Gaussian case.
The parameter $\alpha$ is sometimes called the stability parameter of the stable distribution, and it characterizes the behavior of $p(\rho)$ for large values of $\rho$. 
Because of this relation to stable probability distributions, we shall call the model~\eqref{eq:stable} \textit{stable color-glass condensate} (sCGC) in the following.

\subsection{Dipole amplitude}
As demonstrated in Sec.~\ref{sec:wilson-line_correlators}, to compute the dipole amplitude we only need the function
\begin{equation}
    F^R_z(\xt ,\yt)
    =  \int_0^\infty \dd{\sigma} \xi_R(\sigma) \w_z\qty( \qty[\frac{\gs}{2\pi} K_{\xt \yt \zt} \sigma]^2 )
    = \abs{\frac{\gs}{2\pi} K_{\xt \yt \zt}}^\alpha \mu_z^2 \int_0^\infty \dd{\sigma} \xi_R(\sigma)   \sigma^\alpha .
\end{equation}
Defining
\begin{equation}
    \hat F_R(\alpha) = \int_0^\infty \dd{\sigma} \xi_R(\sigma)   \sigma^\alpha,
\end{equation}
we then see that the representation of the color matrices appears in the function $F^R_z$ only as an overall coefficient.
This coefficient is studied in Appendix~\ref{app:Fhat} in more detail.
We can then write the dipole amplitude as
\begin{equation}
    \label{eq:dipole_stable}
    N(\xt,\yt)
    = 1 - \exp\qty(
    - \hat F_R(\alpha) 
    \int \dd[3]{z}\abs{\frac{\gs}{2\pi} K_{\xt \yt \zt}}^\alpha \mu_z^2 
    ).
\end{equation}

Let us now consider the result Eq.~\eqref{eq:dipole_stable} in more detail.
\update{First of all, we note that for a target with a finite radius $R$, 
the function $\mu_z^2$ describing the color-field strength of the target behaves as $\mu_z^2 \to 0$ for large coordinates $\abs{z} \gg R$.
We can then take the infrared regulator $m $ to zero as the target radius is enough to regulate contributions from large dipoles, finding:
\begin{equation}
K_{\xt \yt \zt} 
\overset{m \to 0}{=}
\log( \frac{\abs{\yt -\zt}}{\abs{\xt -\zt}} ).
\end{equation}
}

\update{
Second, it is interesting to consider the behavior for small dipoles $r=\abs{\xt - \yt}$ such that $r \ll R$.
In this limit, we find:
}
\begin{equation}
\label{eq:small-r}
      \int \dd[3]{z}\abs{\frac{\gs}{2\pi} K_{\xt \yt \zt}}^\alpha \mu_z^2 \propto
    \begin{cases}
        \qty( \frac{\as}{\pi} \frac{r^2}{R^2})^{\alpha/2}  & \text{if } \alpha<2, \\
         \frac{\as}{\pi} \frac{r^2}{R^2} \log(\frac{R}{r})  & \text{if } \alpha=2.
    \end{cases}
\end{equation}
Barring the logarithmic term for $\alpha=2$, we see that the small-$r$ behavior is a power law where the power is determined by the stability parameter $\alpha$.
\update{While this small-$r$ limit is shown for a finite target with $m=0$, the exactly same small-$r$ behavior can be found for an infinite target with a non-zero $m$, effectively replacing $R \to 1/m$ in Eq.~\eqref{eq:small-r}.}
This form is similar to the MV$^\gamma$ model
that has been used in phenomenological fits to the dipole amplitude~\cite{Albacete:2010sy,Lappi:2013zma,Beuf:2020dxl,Casuga:2023dcf,Casuga:2025etc}, missing only the logarithmic term.
While the MV$^\gamma$ model has been motivated by the DGLAP evolution of the gluon density, 
the calculation presented here provides another motivation starting from a modification to the probability distribution for the color-charge density.
This argument can also be turned around to motivate the model~\eqref{eq:stable}, as the power $\alpha < 2$ can be understood as a gluon distribution where the DGLAP evolution has already been taken into account to modify the leading-order behavior $\alpha = 2$ from matching the dipole amplitude to the gluon distribution:
at leading order, one can relate the small-$r$ limit of the dipole amplitude to the gluon parton distribution function (PDF) $g(x,\mu^2)$ as~\cite{Baier:1996sk,Mueller:1999wm,Mueller:2001fv,Kovchegov:2012mbw}
\begin{equation}
\label{eq:N_PDF}
    N_x(\xt ,\yt) \propto r^2  \as  x g(x, 1/r^2).
\end{equation}
In Mellin space, the leading-order DGLAP evolution of the gluon PDF can be solved as
\begin{equation}
    \hat g(\omega,\mu) \sim \qty(\mu^2)^{\as/(2\pi) \times \gamma_{gg}(\omega)}
\end{equation}
where $\gamma_{gg}$ is the anomalous dimension for the gluon--gluon splitting and we have ignored the quark contribution.
The $\mu$-dependence of the gluon PDF then suggests a modification of the small-$r$ behavior of the dipole amplitude from $r^2$ to $r^\alpha$ with $\alpha<2$. 
Similar corrections to the power law can also be found from transverse broadening in heavy nuclei~\cite{Caucal:2022fhc}.
These computations suggest that the modifications to the power $\alpha$ should be perturbatively calculable, and it would be interesting to study the relation~\eqref{eq:N_PDF} at higher orders to determine how the DGLAP evolution modifies the small-$r$ behavior exactly.

The sCGC model also explains why the MV$^\gamma$ model with parameter values $\gamma = \alpha/2 > 1$ is found to be problematic as it leads to a negative gluon distribution~\cite{Lappi:2013zma, Giraud:2016lgg, Shi:2021hwx, Casuga:2023dcf,Casuga:2025etc}:
for values $\alpha > 2$, the probability distribution $p[\rho]$ is guaranteed to have negative values, meaning that it cannot be used as a valid probability distribution to model the local color-charge distribution.
Thus, problematic values $\alpha > 2$ are excluded automatically in this model.

\subsection{\texorpdfstring{Large-$\nc$ limit}{Large-Nc limit}
}

It is also interesting to study the large-$\nc$ limit in more detail.
Especially, we wish to see if the mean-field limit
\begin{equation}
    \label{eq:mean_field}
    \expval{ S^{(2)}(\xt,\yt) S^{(2)}(\yt,\xt)} \approx \expval{ S^{(2)}(\xt,\yt)} \times  \expval{   S^{(2)}(\yt,\xt)}
\end{equation} 
is exact at large $\nc$ in the sCGC model.
The left-hand side of this is given by 
\begin{equation}
\label{eq:double_dipole}
  \expval{ S^{(2)}(\xt,\yt) S^{(2)}(\yt,\xt) }
    \overset{\nc \to \infty}{=} \exp\qty(
    - \hat F_A(\alpha) 
    \int \dd[3]{z}\abs{\frac{\gs}{2\pi} K_{\xt \yt \zt}}^\alpha \mu_z^2 
    )
\end{equation}
whereas the right-hand side is
\begin{equation}
\expval{ S^{(2)}(\xt,\yt)} \times  \expval{   S^{(2)}(\yt,\xt)}
    \overset{\nc \to \infty}{=}
    \exp\qty(
    - 2\hat F_F(\alpha) 
    \int \dd[3]{z}\abs{\frac{\gs}{2\pi} K_{\xt \yt \zt}}^\alpha \mu_z^2 
    ).
\end{equation}
The validity of the mean-field limit then corresponds to the identity
\begin{equation}
\label{eq:limit_equivalence}
    \hat F_A(\alpha) \overset{?}{=} 2 \hat F_F(\alpha).
\end{equation}
By checking the functional forms of $\hat F_R$ for the representations $\su_F(\infty)$ and $\su_A(\infty)$ from Table~\ref{tab:F_hat} in Appendix~\ref{app:Fhat}, we see that this equation is in general \textit{not} valid.
From the allowable values $0 < \alpha \leq 2$, it is only valid for $\alpha =2$ which is the standard Gaussian case.
Thus, we see that the mean-field limit is generally not satisfied even in the large-$\nc$ limit, and these two limits correspond to different approximations.
The difference between these two limits can be understood to come from the heavy tail of the stable distribution. 
This means that the gluon field can have very large values with a significant probability, and thus higher-order correlators between the gluon fields that could normally be neglected in the large-$\nc$ limit need to be taken into account.

The heavy tail of the stable distribution also
has some surprising ramifications when considering the dilute limit where the Wilson-line correlators are expanded to the first non-trivial order in $\mu_z^2$.
Notably, in the dilute limit the Wilson-line structure of the JIMWLK evolution for the dipole amplitude,
\begin{equation}
    W_\text{JIMWLK}(\xt, \yt, \zt)
    = 
    \expval{
    S^{(2)}(\xt,\yt) - S^{(2)}(\xt,\zt) S^{(2)}(\zt,\yt)
    },
\end{equation}
is expected to reduce to the BFKL one:
\begin{equation}
    W_\text{BFKL}(\xt, \yt, \zt)
    = 
    \expval{
    S^{(2)}(\xt,\yt) + 1 - S^{(2)}(\xt,\zt) - S^{(2)}(\zt,\yt)
    }
    = N(\xt,\zt) + N(\zt,\yt) - N(\xt,\yt).
\end{equation}
Let us study what happens in the limit $\xt = \yt$.
Then we can use Eq.~\eqref{eq:4point} to simplify the JIMWLK expression as
\begin{equation}
    W_\text{JIMWLK}(\xt, \xt, \zt)
    = 
    \expval{
    1 - S^{(2)}(\xt,\zt) S^{(2)}(\zt,\xt)
    }
    \overset{\text{dilute limit}}{\approx}
    \frac{\nc^2-1}{\nc^2}
     \hat F_A(\alpha) 
    \int \dd[3]{z}\abs{\frac{\gs}{2\pi} K_{\xt \yt \zt}}^\alpha \mu_z^2,
\end{equation}
and correspondingly for the BFKL case we have
\begin{equation}
    W_\text{BFKL}(\xt, \xt, \zt)
    = 
    \expval{
    2 - S^{(2)}(\xt,\zt) - S^{(2)}(\zt,\xt)
    }
      \overset{\text{dilute limit}}{\approx}
     2 \hat F_F(\alpha) 
    \int \dd[3]{z}\abs{\frac{\gs}{2\pi} K_{\xt \yt \zt}}^\alpha \mu_z^2.
\end{equation}
Similarly to Eq.~\eqref{eq:limit_equivalence}, these equations are equal  only for the Gaussian case $\alpha=2$ and otherwise they differ by a constant factor.
While this observation may initially strike as odd, it follows from the fact that in the sCGC approximation the target is never truly dilute for $\alpha < 2$ due to the heavy tails of the color-charge distribution.
These heavy tails then correspond to strong color-charge distributions that are probable enough so that they cannot be neglected.
For this reason, there is no a priori reason to expect the dilute approximation to work for the sCGC model, in this case invalidating the derivation of the BFKL evolution that relies on being able to neglect scatterings of multiple BFKL ladders.

\subsection{Numerical implementation}

While we have now shown how to compute practically any physical Wilson-line correlator analytically with the weight function~\eqref{eq:W_general}, in some cases one needs a numerical implementation for evaluating the correlators.
This requires sampling the radial part of the probability distribution $p\qty( \rhosq)$ which can be non-trivial.
We will now show how to do this in practice for the sCGC model.

The probability density for the sCGC model is defined as 
\begin{equation}
\label{eq:p_stable}
    p\qty(\rhosq)
    = \int \dd[D]{\vec \sigma} \qty(\frac{\DD[3]{z}}{2\pi })^D \exp[-i \DD[3]{z} \vec \sigma \vdot \vec \rho]  
    \exp( - \DD[3]{z} \mu^2 \sigma^\alpha )
\end{equation}
which does not generally have an existing numerical implementation.
However, we can use Point~\ref{point:power_law} discussed at the beginning of this section to sample this distribution:
\textit{any} probability distribution with a single scale parameter produces the same results for the Wilson-line correlators as the sCGC probability distribution~\eqref{eq:p_stable} for some parameter values $\mu^2$ and $\alpha$.

Consider now that we want to sample the sCGC distribution $p\qty( \rhosq)$ using a probability distribution $p'(\rho)$ that only depends on $\rho^2 = \rhosq$.
To find the corresponding values of $\mu^2$ and $\alpha$, we first note that
\begin{equation}
    \int \dd[D]{\vec \rho} e^{i \DD[3]{z} \vec \rho \vdot \vec \sigma}  p\qty(\rhosq)
    =  \exp( - \DD[3]{z} \mu^2 \sigma^\alpha )
    \approx 1 - \DD[3]{z} \mu^2 \sigma^\alpha + \mathcal{O}\qty( \qty[\DD{z^+}]^2 ).
\end{equation}
On the other hand, for any probability distribution $p'(\rho)$ we can compute
\begin{equation}
\label{eq:pprime_expansion}
\begin{split}
    \int \dd[D]{\vec \rho} e^{i \DD[3]{z} \vec \rho \vdot \vec \sigma}  p'(\rho) p_\Omega(\Omega_\rho)
    =& \frac{2\pi}{\vol{S^{D-1}}} \int_0^\infty \dd{\rho} 
    \qty( \frac{2\pi}{\DD[3]{z} \sigma \rho})^{D/2 - 1} J_{D/2 - 1}\qty( \DD[3]{z} \sigma \rho) p'(\rho)\\
    =& 1 - C \sigma^\beta + \ldots
\end{split}
\end{equation} 
where $p_\Omega(\Omega_\rho) = 1 / \vol{S^{D-1}}$ is the angular part of the probability distribution, and in the final line we have expanded in $\sigma$ and kept only the leading non-trivial term.
We can then match the parameters in $p'(\rho)$ by demanding that 
$C = \DD[3]{z} \mu^2 $ and $\beta = \alpha$, 
allowing us to use $p'(\rho)$ to sample the sCGC distribution $p\qty(\rhosq)$.
The mismatch between the two probability distributions corresponds to terms of higher order in $\DD{z^+}$ which are not relevant for computing Wilson-line correlators.

As a practical example, we list some probability distributions for $\rho \in [0,\infty)$ that can be used for sampling the sCGC distribution:
\begin{enumerate}
 \item Scaled beta prime distribution:
 \begin{equation}
     p'(\rho; C, \gamma, \beta ) =\frac{C^\gamma}{\text{B}(\gamma,\beta)} \rho^{\gamma-1} (1+C\rho)^{-\gamma -\beta}
 \end{equation}
 where $\text{B}(\gamma, \beta)$ is the beta function.
 The parameter values are given by
 \begin{align}
     \beta &= \alpha,
     &
     C &= \qty[
     \qty( \DD[3]{z} )^{1-\alpha}
     \times \mu^2 
     \times
     \frac{ 2^{\alpha}
     \Gamma(1 + \alpha)\Gamma(\gamma)
     \Gamma\qty( \frac{\alpha + D}{2} )
     }{
     \Gamma\qty(1 -\frac{\alpha }{2} )
        \Gamma\qty(\alpha + \gamma )
     \Gamma\qty( \frac{D}{2} )
     }
     ]^{-1 / \alpha},
 \end{align}
 and $\gamma >0$ is arbitrary.
 \item Inverse gamma distribution:
 \begin{equation}
       p'(\rho; C, \gamma )
       =
       \frac{C^\gamma}{\Gamma(\gamma)} \frac{e^{-C/\rho}}{\rho^{\gamma+1}}
 \end{equation}
 with
 \begin{align}
     \gamma &= \alpha,
     &
     C &= \qty[
     \qty( \DD[3]{z} )^{1-\alpha}
     \times \mu^2 
     \times
     \frac{ 2^{\alpha}
     \Gamma(1 + \alpha)
     \Gamma\qty( \frac{\alpha + D}{2} )
     }{
     \Gamma\qty(1 -\frac{\alpha }{2} )
     \Gamma\qty( \frac{D}{2} )
     }
     ]^{1 / \alpha}.
 \end{align}
 \item Scaled one-sided Student's $t$ distribution:
 \begin{equation}
     p'(\rho; C, \nu )
     = 2C \times
     \frac{ \Gamma\qty(\frac{\nu +1}{2})}{\sqrt{\pi \nu}\Gamma\qty(\frac{\nu }{2}) }
     \qty( 1 + \frac{C^2 \rho^2}{\nu} )^{-(\nu+1)/2}
 \end{equation}
 with
 \begin{align}
     \nu &= \alpha,
     &
     C &= \qty[
     \qty( \DD[3]{z} )^{1-\alpha}
     \times \mu^2 
     \times
     \qty(\frac{2}{\sqrt{\alpha}})^{\alpha}
     \times
     \frac{
     \Gamma\qty( \frac{1}{2})
     \Gamma\qty(1 + \frac{\alpha}{2})
     \Gamma\qty( \frac{\alpha + D}{2} )
     }{
     \Gamma\qty(\frac{1+\alpha }{2} )
     \Gamma\qty(1 -\frac{\alpha }{2} )
     \Gamma\qty( \frac{D}{2} )
     }
     ]^{-1 / \alpha}.
 \end{align}
    \item One-dimensional one-sided stable distribution:
    \begin{equation}
        p'(\rho; C, \gamma ) = 2 \int_{-\infty}^\infty \frac{\dd{\sigma}}{2\pi}
        e^{- i\rho \sigma - \abs{C \sigma}^{\gamma} }
    \end{equation}
    with
    \begin{align}
        \gamma &= \alpha,
        &
        C &= \qty[ \qty( \DD[3]{z} )^{1-\alpha}
        \times
        \mu^2 
        \times \frac{\Gamma(\frac{1}{2})\Gamma(\frac{\alpha + D}{2})}{\Gamma(\frac{1 + \alpha}{2})\Gamma(\frac{D}{2})} ]^{1/\alpha}.
    \end{align}
\end{enumerate}
The last two distributions are usually defined on the whole real axis $(-\infty, \infty)$ instead of the semi-infinite interval $[0,\infty)$ required for $\rho = \sqrt{ \rhosq }$, but in practice one can simply take the absolute value of the sampled values to get only positive-valued samples.

In general, these distributions work for values $0 < \alpha < 2$, with the exception that the stable distribution can also be used to sample the Gaussian case $\alpha =2$.
The reason for this is that for the other distributions the subleading term in the expansion in Eq.~\eqref{eq:pprime_expansion}
is of the order $\mathcal{O}(\sigma^2)$, which would need to be taken into account for $\alpha =2$.
For the stable distribution, the next term in the expansion would be $\mathcal{O}(\sigma^{2\alpha})$ instead, allowing us to extend it to the Gaussian case.
Considering the order of the subleading term also allows us to estimate the error in terms of the lattice spacing $\DD{z^+}$ when using these distributions for sampling:
for the stable distribution, we estimate the error to be 
$\mathcal{O}\qty(\qty[\DD{z^+}]^{\alpha})
$, while for the other  distributions the error is
$\mathcal{O}\qty(\qty[\DD{z^+}]^{2/\alpha - 1})
$.

\begin{figure}
	\centering
    \begin{subfigure}{0.45\textwidth}
        \centering
        \includegraphics[width=\textwidth]{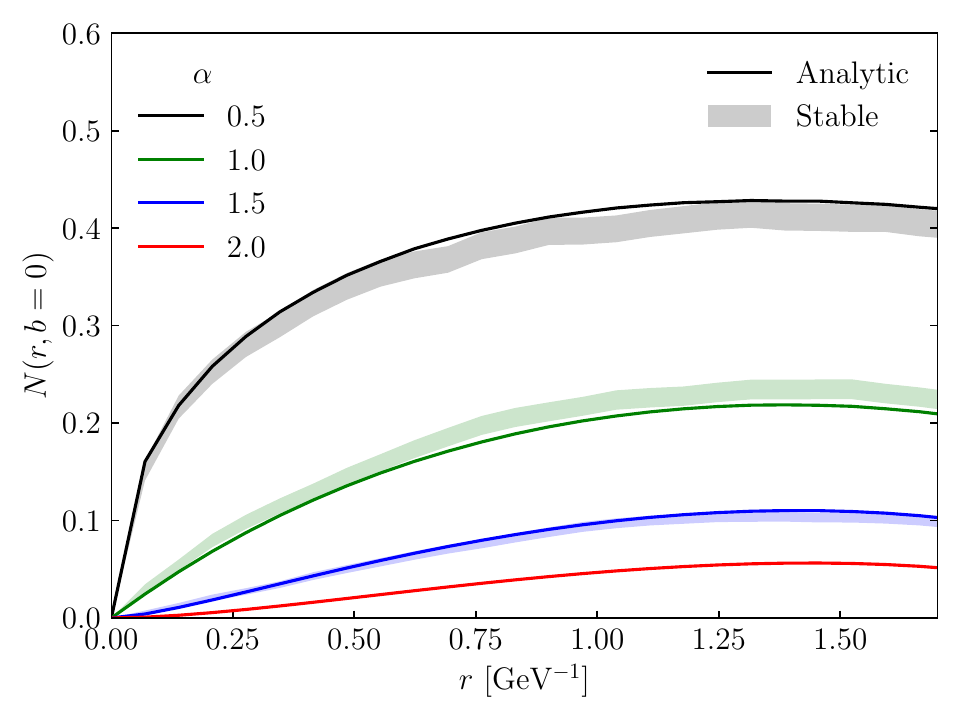}
        \caption{ Stable distribution.}
        \label{fig:stable}
    \end{subfigure}
    \begin{subfigure}{0.45\textwidth}
        \centering
        \includegraphics[width=\textwidth]{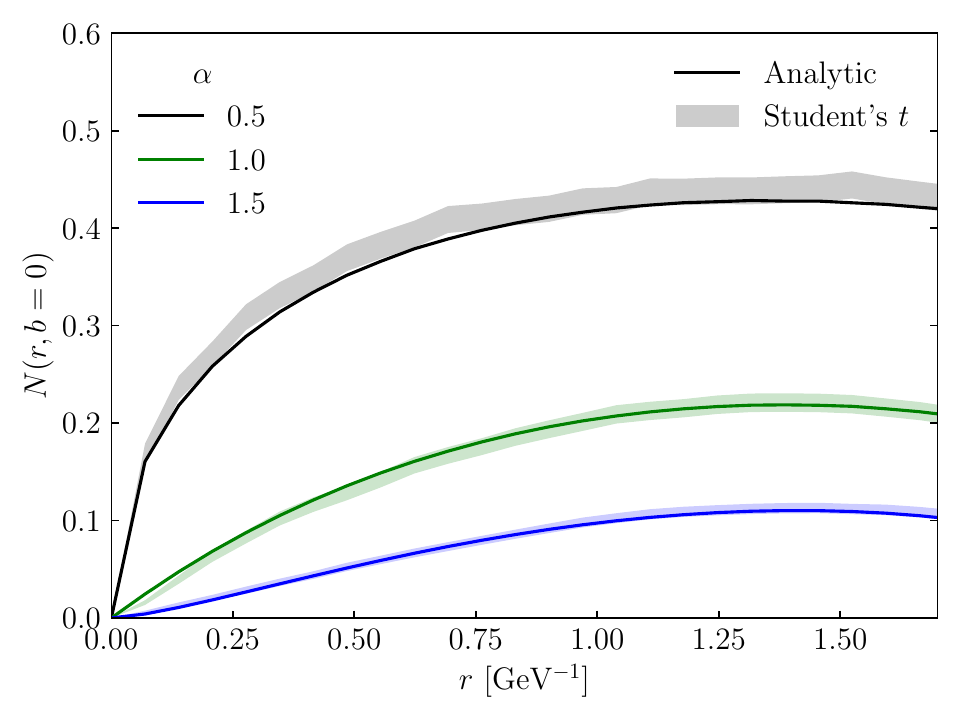}
        \caption{ Student's $t$ distribution.}
        \label{fig:t}
    \end{subfigure}
    
    \begin{subfigure}{0.45\textwidth}
        \centering
        \includegraphics[width=\textwidth]{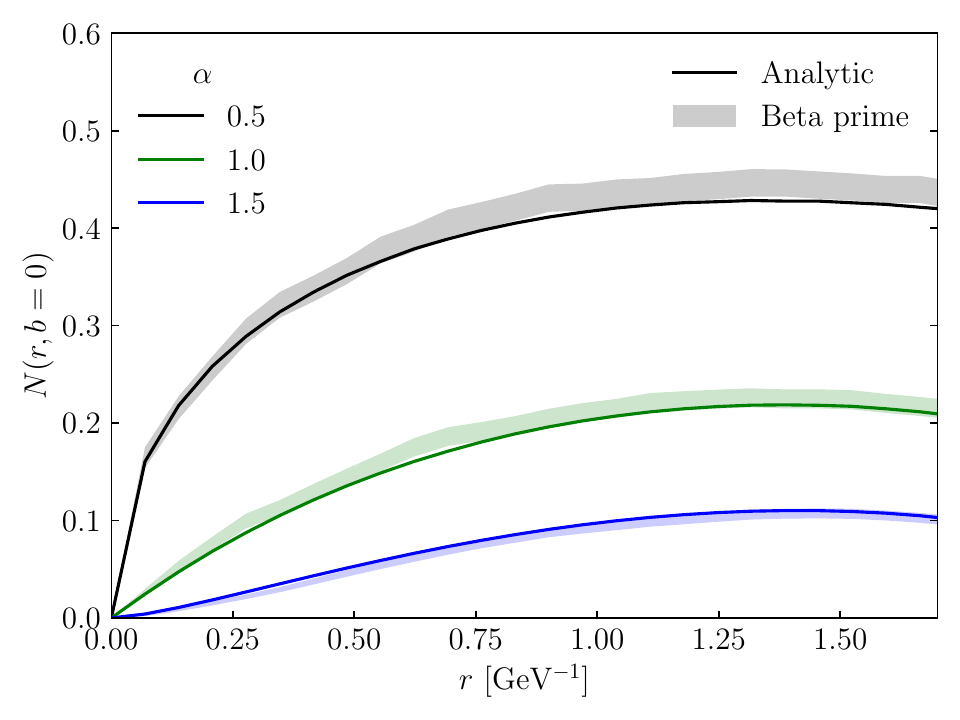}
        \caption{ Beta prime distribution.}
        \label{fig:beta_prime}
    \end{subfigure}
    \begin{subfigure}{0.45\textwidth}
        \centering
        \includegraphics[width=\textwidth]{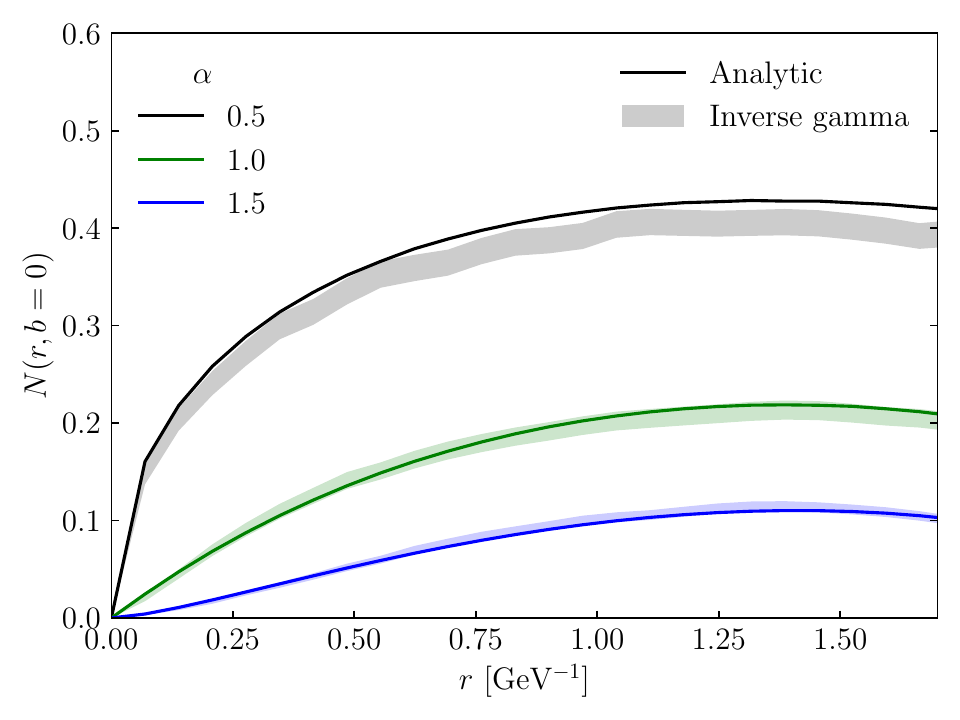}
         \caption{ Inverse gamma distribution.}
        \label{fig:inverse_gamma}
    \end{subfigure}
	  \caption{The dipole amplitude in the sCGC model evaluated as a function of the dipole size $r = \abs{\xt -  \yt}$ at zero impact parameter $b = \frac{1}{2} \abs{\xt + \yt} = 0$.
      Different figures correspond to using different probability distributions for numerical sampling.
      The analytical results are shown as solid lines, whereas the shaded bands correspond to the $1\sigma$ variation of the numerically sampled results.
      }
        \label{fig:scgc}
\end{figure}

As an illustration of the numerical implementation, we have checked the analytical formula for the dipole amplitude~\eqref{eq:dipole_stable} numerically.
This is shown in Fig.~\ref{fig:scgc}, where we have plotted the dipole amplitude in the sCGC model using different probability distributions for sampling.
For the beta prime distribution, we have chosen $\gamma=1$.
The results have been computed using a finite lattice where each of the lattice points has its own probability distribution.
The exact values of the parameters used in making these plots are 
the following:
\begin{itemize}
    \item Number of colors: $\nc = 3$,
    \item Scale for the target size: $B = \SI{3}{GeV^{-2}}$,
    \item Transverse width of the lattice:
    $w_T = \sqrt{B}$,
    \item Longitudinal width of the lattice:
    $w_L = \sqrt{B}$,
    \item Number of transverse lattice points:
    $N_T = 51$,
    \item Number of longitudinal  lattice points:
    $N_L = 100$,
    \item Transverse lattice spacing:
    $\DD[2]{\zt} = (w_T / N_T)^2$,
    \item Longitudinal lattice spacing:
    $\DD{z^+} = w_L / N_L$,
    \item Number of samples:
    $N_\text{sample}= \num{1000}$,
    \item Infrared regulator: $m = \SI{0.1}{GeV}$,
    \item Color-field strength:
    \begin{equation}
        \mu_z^2 = \kappa \times \frac{1}{2\pi B^{3/2}} \exp( -\frac{\zt^2}{2B}  )
    \end{equation}
    where $\kappa= 10$,
    \item Coupling constant: $g_s = 1$.
\end{itemize}
We note that for actual phenomenological studies the lattice width and the number of lattice points are too small, and finite-lattice effects are expected to be large.
However, as long as the analytical expression~\eqref{eq:dipole_stable} is computed using the same lattice, the analytical results should agree with the numerical ones.
Indeed, this is what we see in Fig.~\ref{fig:scgc},
where the results agree within the sampling uncertainty.
The sampling uncertainty is estimated from the $1\sigma$ standard deviation of the sample mean, which is generally at the $5\%$ level for all distributions.
One exception for this is the Gaussian case $\alpha=2$ for which the error is much smaller, of the order $1 \%$.
This is expected as the Gaussian probability distribution does not have heavy tails unlike the other distributions, leading to a better convergence when averaging over the color-field configurations.
The sampling uncertainty could be reduced further by increasing the number of samples.

\section{Discussion}
\label{sec:discussion}

In this work, we have considered an extension of the standard color-glass condensate model where the weight function is assumed to be Gaussian.
Our model works for a wide class of weights where the color distributions at different coordinates are assumed to be uncorrelated.
We have shown how to compute general Wilson-line correlators in this model by using a differential equation that is analogous to the ones used in deriving higher-order correlators within the Gaussian approximation. 
For the dipole amplitude,
the analytical form obtained is very simple,
which makes the implementation of these types of models fairly straightforward for actual phenomenological studies of the target structure in the small-$x$ limit.

In particular, we have proposed the sCGC model based on stable probability distributions,
which extends the Gaussian case naturally to distributions that satisfy a generalized version of the central limit theorem.
Additionally, it turns out that any local probability distribution that is described by a single scale parameter will revert to the sCGC model in the continuum limit for some stability parameter $\alpha$.
This stability parameter also controls the behavior of the dipole amplitude for small dipole sizes, changing it from the Gaussian case $N \sim r^2 \log r$ to a power law $N \sim r^\alpha$.
Such a modification of the small-$r$ behavior could allow for an approximate inclusion of the DGLAP evolution to the dipole amplitude, as the value of the power $\alpha$ controls the dependence of the cross section on the momentum scale.
These observations suggest that the sCGC model might be able to capture most of the relevant information of the target structure for high-energy scattering.

We have also checked
our analytical results for the sCGC model numerically.
This has been done by numerically evaluating the color-field average,
and we have provided examples of how to do this in practice.
The sCGC model can then be used in phenomenological applications where higher-order correlators are also needed, and in cases that require sampling individual Wilson lines of the correlators such as the IP-Glasma model~\cite{Schenke:2012wb,Schenke:2012hg,Gale:2012rq,Schenke:2013dpa} in heavy-ion collisions and the JIMWLK evolution~\cite{Jalilian-Marian:1997qno,Jalilian-Marian:1997ubg,Kovner:2000pt,Iancu:2000hn,Mueller:2001uk} in its Langevin form~\cite{Weigert:2000gi,Blaizot:2002np,Dumitru:2011vk,Lappi:2012vw,Cali:2021tsh}.

\update{
One interesting addition to the model would be considering correlations in the color charges at different light-cone coordinates.
Even though these correlations are not expected to be important for large nuclei, where the individual nucleons act as color sources,
such correlations could be necessary for proton targets.
However, while}
the color weights considered in this work still ignore correlations between different coordinates, 
this class of models provides a very flexible parametrization for the nuclear color structure that can be tested against experimental data.
With the precise data expected to be collected at the upcoming Electron--Ion Collider~\cite{Accardi:2012qut,Aschenauer:2017jsk,AbdulKhalek:2021gbh}, it should be possible to study the fine details of nuclear structure in the high-energy limit.

\begin{acknowledgments}
I would like to thank Paul Caucal, Adrian Dumitru, Yoshitaka Hatta, Tuomas Lappi, Heikki Mäntysaari, Yacine Mehtar-Tani, Farid Salazar, Björn Schenke, and Raju Venugopalan for useful discussions and comments.
The author is supported by the National Science Foundation under grant No. PHY-1945471, and by the U.S. Department of Energy, Office of Science, Office of Nuclear Physics, within the framework of the Saturated Glue (SURGE) Topical Theory Collaboration.
The author is also grateful to the Mani L. Bhaumik
Institute for Theoretical Physics for support.
\end{acknowledgments}

\appendix

\section{\texorpdfstring{Computing $\xi_R$ for a general representation of $\su(\nc)$}{Computing xiR}
}
\label{app:moment-generating_function}

\subsection{\texorpdfstring{General approach for finite $\nc$}{General approach for finite Nc}
}

In this Appendix, we wish to review the evaluation of the function $\xi_R(\sigma)$
in Eq.~\eqref{eq:xi}.
The main difficulty in computing this function comes from the integration over the angle $\Omega_\rho$, as the exponential matrix is a nontrivial function of this angle.
There is, however, a different way of thinking about this integral that allows us to compute it for a general representation.
The main idea is to recognize that the trace we wish to evaluate can be identified with the \textit{character}
\begin{equation}
    \chi_R(g) = \Tr[R(g)]
\end{equation}
mapping elements of a group to complex numbers.
Here $g$ is an element of the group and $R(g)$ is the corresponding matrix in the representation $R$.
It turns out that the value of the character depends only on the \textit{conjugacy class} of the group element $g$, and these classes can be labeled by the eigenvalues of the matrices in the fundamental representation.
Specifically, we will consider eigenvalues $\vec \lambda = (\lambda_1 , \ldots , \lambda_{\nc})$ of the matrices $\hat \rho^a t^a$ in the fundamental representation where $\hat \rho^a$ is a unit vector.
Conjugacy classes do not depend on the representation, and thus we can use
these same eigenvalues to label the class in any representation, meaning that the value of the character $\chi_R$ is completely determined by $\vec \lambda$.
If we can then change the integration over the angle $\Omega_\rho$ to integration over the eigenvalues $\vec \lambda$ and compute the character function, we can determine the value of $\xi_R(\sigma)$.

Let us first consider how to compute the character $\chi_R$.
For the group $\su(\nc)$, the Weyl character formula implies that we can write the character as~\cite{MR255,Chen:2002gd,Greiner:1989eu,MR1153249}
\begin{equation}
    \chi_R(\vec \lambda) \equiv \Tr[ \exp\qty( i t_R(\vec \lambda) ) ]
   =  s_\mu\qty(\vec \lambda)
\end{equation}
where
$s_\mu$ is the Schur polynomial of the partition $\mu$.
The relevant partition $\mu$ of the representation can be determined from the \textit{Dynkin label} $\vec \alpha  = (\alpha_1, \ldots, \alpha_{\nc - 1})$ of the representation, where $\alpha_n$ counts the number of columns of length $n$ in the corresponding Young tableau~\cite{Chen:2002gd}.
Specifically, if the representation has the Dynkin label $\vec \alpha$, then the corresponding partition $\mu = (\mu_1, \ldots , \mu_{\nc})$ is given by~\cite{Chen:2002gd,Greiner:1989eu}
\begin{equation}
\mu_n =
    \begin{cases}
       0 & \text{ if } n = \nc,\\
       \sum_{i=n}^{\nc-1} \alpha_i & \text{ otherwise}. 
    \end{cases}
\end{equation}
Some examples of the Dynkin labels $\vec \alpha$ and partitions $\mu$ are shown in Table~\ref{tab:reps}.

\begin{table}[t]
  \renewcommand{\arraystretch}{2}
    \centering\begin{tabular}{cccc}
\toprule 
 Representation & Dynkin label $\vec \alpha$   & Partition $\mu$ & Conjugate partition $\nu$
\\ \midrule
$F$ & $(1,0, \ldots, 0)$  & $(1,0, \ldots ,0)$ & $ (1) $
\\
$\overline F$ & $(0, \ldots, 0, 1)$  & $(1,1, \ldots ,1,0)$ & $ (\nc -1) $
\\
$A$ & $(1,0, \ldots, 0, 1)$  & $(2,1, \ldots ,1, 0)$ & $ (\nc -1, 1) $
\\
\bottomrule
\end{tabular}
    \caption{
    The Dynkin label $\vec \alpha$, the corresponding partition $\mu$, and the conjugate partition $\nu$ for various representations of $\su(\nc)$.
    Here $F$, $\overline F$, and $A$ stand for fundamental, antifundamental, and adjoint representations.
    }
    \label{tab:reps}
\end{table}

To actually compute the Schur polynomials in our case, it is easiest to use the second Jacobi--Trudi formula that relates the Schur polynomial $s_\mu$ to \textit{elementary symmetric polynomials} $e_{n}$~\cite{MR1153249}:
\begin{equation}
     s_\mu\qty(\vec \lambda) = 
     \mdet{
        e_{\nu_1}        & e_{\nu_1 + 1 }   &\cdots & e_{\nu_1 + l -1} \\
        e_{\nu_2-1}      & e_{\nu_2 }       &\cdots &e_{\nu_2 + l -2} \\
       \vdots            & \vdots           &\ddots &\vdots \\
        e_{\nu_l - l +1} & e_{\nu_l -2 +1 } &\cdots &e_{\nu_l } 
     }
\end{equation}
where $\nu = (\nu_1, \ldots , \nu_l)$ is the conjugate partition of $\mu$ (see Table~\ref{tab:reps} for the conjugate partitions of the relevant representations).
The elementary symmetric polynomials $e_n$ are defined as
\begin{equation}
    e_n = \sum_{1 \leq j_1 < j_2 < \cdots < j_n \leq \nc} \Lambda_{j_1} \cdots \Lambda_{j_n}
\end{equation}
where
$\Lambda_i = \exp( i \lambda_i )$.
This leads to the following expressions for the characters of the fundamental ($F$), antifundamental ($\overline F$), and adjoint ($A$) representations:
\begin{align}
\label{eq:character_F}
\chi_F(\vec \lambda)  &= 
e_1 
&
&=
\sum_{i=1}^{\nc} \Lambda_i  
&
&= \sum_{i=1}^{\nc} \exp[ i \lambda_i ] 
      \\
\label{eq:character_antiF}
\chi_{\overline F}(\vec \lambda)  &= 
e_{\nc - 1}
&
&=
\sum_{i=1}^{\nc} \Lambda_i^{-1}  
&
&= \sum_{i=1}^{\nc} \exp[ - i \lambda_i ] 
      \\
\label{eq:character_A}
      \chi_A(\vec \lambda)  &=
      \mdet{
      e_{\nc - 1} & e_{\nc} \\
      e_{0}       & e_1
      }
      &
      &=
      -1 + \qty[\sum_{i=1}^{\nc} \Lambda_i]  \qty[\sum_{j=1}^{\nc} \Lambda_j^{-1}   ]
     &
     &= -1 + \sum_{i=1}^{\nc}  \sum_{j=1}^{\nc} \exp[ i (\lambda_i-\lambda_j) ]
\end{align}
where we have also used the fact that for $\su(\nc)$ we have $ \sum_{i=1}^{\nc} {\lambda_i}  = 0 $
and thus $ \prod_{i=1}^{\nc} {\Lambda_i}  = 1 $.

We still need to understand how to transform the integral over the angle $\Omega_\rho$ to an integral over the eigenvalue vector $\vec \lambda$.
This corresponds to finding the probability distribution for a uniformly distributed vector $\vec \lambda$.
The uniform probability density
is given by the Haar measure of the group, and for unitary matrices U($\nc$) this can be written as~\cite{MR255}
\begin{equation}
\label{eq:pU}
    p_\text{U}(\vec \lambda)
    \propto \prod_{i < j} \abs{\Lambda_i - \Lambda_j}^2
\end{equation}
up to a normalization constant.
We can then use this to determine the probability distribution for the group $\su(\nc)$.
This can be done by the following trick:
First, we note that we are only interested in the distribution for the angular part of the eigenvalues $\vec \lambda$ 
which cannot depend on the normalization of the eigenvalues.
Second, we note that in the limit $\vec \lambda^2 \ll 1$ the relation between the eigenvalues $\lambda_i$ and the group elements $\Lambda_i = \exp(i \lambda_i) \approx 1 + i \lambda_i$ is linear, meaning that their probability distributions have to agree in this limit.
We can then expand the distribution~\eqref{eq:pU} for small $\vec \lambda$:
\begin{equation}
    p_{\ua}(\Omega_\lambda) \propto \lim_{\vec \lambda^2 \to 0}  \prod_{i < j} \abs{\Lambda_i - \Lambda_j}^2
    \propto \prod_{i < j} (\lambda_i - \lambda_j)^2
\end{equation}
where 
this probability distribution is understood to depend only on $\Omega_\lambda$ which is the angle of the vector $\vec \lambda$;
note that the overall normalization of the vector $\vec \lambda$ appears as an overall constant that will cancel when the probability distribution is properly normalized.
We can further restrict this probability distribution to $\su(\nc)$ by introducing a delta function that enforces $\sum_i \lambda_i = 0$.
We then find
\begin{equation}
     p_{\sua}(\Omega_\lambda) \propto
     \delta\qty( \sum_i\lambda_i)
     \prod_{i < j} (\lambda_i - \lambda_j)^2
\end{equation}
up to a normalization constant.
This result can be simplified even further by noting that this delta function also has a geometrical interpretation~\cite{MR3439116}:
it says that the vectors $\vec \lambda$ and $(1,\ldots,1)$ and are orthogonal. 
This means that instead of thinking about the vector $\vec\lambda$ on the sphere $S^{\nc - 1}$ (as we only care about the angular part and not the overall normalization),  we only need to consider the projection of the vector $\vec \lambda$ to the sphere $S^{\nc-2}$ that is orthogonal  to the vector $(1,\ldots,1)$.
We can then replace the integration over the angle $\Omega_\rho$ with
\begin{equation}
\label{eq:gamma}
\begin{split}
    \gamma_R( \rho )
    =&
    \int_{S^{D-1}} 
    \frac{\dd{\Omega_\rho}}{\vol{ S^{D-1} }}
    \frac{1}{D_R}
    \chi_R\qty[\rho \hat \rho^a t_R^a ] \\
    =& \mathcal{N}  \int_{S^{\nc -2}}
    \frac{\dd{\omega_\lambda} }{\vol{ S^{\nc-2} }}
    \frac{1}{D_R}  \chi_R\qty[ \rho \sqrt{T_R} \hat \lambda(\omega_\lambda) ]
       \prod_{i < j} \qty[ \hat \lambda_i(\omega_\lambda) -  \hat \lambda_j(\omega_\lambda)]^2
\end{split}
\end{equation}
where $\hat \lambda$ is a unit vector in the direction of $\vec \lambda$,
$\omega_\lambda$ is the solid angle on the  sphere $S^{\nc -2}$ defining $\hat \lambda$, and 
$T_R$ depends on the representation as
\begin{equation}
    \Tr[t_R^a t_R^a] = T_R D_R.
\end{equation}
The normalization constant $\mathcal{N}$ can be computed from the condition
\begin{equation}
\label{eq:gamma_normlization}
  \gamma_R(0)=  \mathcal{N}  \int_{S^{\nc -2}}
    \frac{\dd{\omega_\lambda} }{\vol{ S^{\nc-2} }}
       \prod_{i < j} \qty[ \hat \lambda_i(\omega_\lambda) -  \hat \lambda_j(\omega_\lambda)]^2
       = 1.
\end{equation}
Here the constant $\vol{ S^{\nc-2} }$  is merely a convention that could also be absorbed into the definition of $\mathcal{N}$.

Finally,
the function $\gamma_R$ can be used to compute $\xi_R$ in Eq.~\eqref{eq:xi} as we note that
\begin{equation}
    \xi_R(\sigma) =
    \vol{ S^{D-1} }
    \int_0^\infty \dd{\rho}  \qty(\frac{\sigma \rho}{2\pi})^{\frac{D}{2}}
    J_{D/2-1} \qty(\sigma \rho) \gamma_R(\rho).
\end{equation}
In general, this integral is defined only in the distributional sense.
In the following examples, it turns out that we can write $\gamma_R(\sigma)$ in terms of Bessel functions, and focusing on the case $\sigma \geq 0$ we can compute this integral using the identity~\eqref{eq:bessel_integral_evaluated} from Appendix~\ref{sec:bessel}.

In practice, computing these functions can be very demanding when the number of colors $\nc$ is large, and
for this reason we will only show the results for the groups $\su(2)$ and $\su(3)$.
The results for the large-$\nc$ limit will be derived using a different method.

\subsection{SU(2)}
\label{sec:su2}

While the case SU(2) is not physically interesting for QCD, it will serve as a demonstration of the general method.
Additionally, we can compute the results for SU(2) directly without using the probability distribution, and for this reason it works as a simple crosscheck of the more general results.

\subsubsection{Fundamental representation  }

In the fundamental representation of SU(2) the calculation is very straightforward.
The color matrices are given by $t^i = \frac{1}{2} \sigma^i$ where $\sigma^i$ are the standard Pauli matrices.
Using the result
\begin{equation}
    \exp\qty( i  \frac{1}{2}\rho^i \sigma^i ) = \cos\qty( \frac{\rho}{2}) + i \hat \rho^i \sigma^i \sin\qty(\frac{\rho}{2}),
\end{equation}
we see that in this case
\begin{equation}
    \gamma_F(\rho) = \int_{S^2} 
    \frac{\dd{\Omega_\rho}}{4\pi} 
    \frac{1}{2} \Tr\qty[   \exp\qty( i  \frac{1}{2}\rho^i \sigma^i ) ]
    =  \cos\qty( \frac{\rho}{2})
\end{equation}
and
\begin{equation}
    \xi_F(\sigma) = 4\pi \int_0^\infty \dd{\rho} \qty(\frac{\sigma \rho}{2\pi})^{3/2} J_{1/2}(\sigma \rho) \times   \cos\qty( \frac{\rho}{2})
    = -\sigma \frac{\dd}{\dd{\sigma}} 
\delta\qty(\sigma-\frac{1}{2}) 
\end{equation}
where 
the integral has been computed by means of Appendix~\ref{sec:bessel} by noting that 
\begin{equation}
    J_{-1/2}(x) =   \sqrt{ \frac{2}{\pi}} \frac{\cos(x)}{\sqrt{x}}.
\end{equation}

\subsubsection{Adjoint representation }
The generators of the adjoint representation of SU(2) are defined as
\begin{equation}
    T^i_{jk} = -i\varepsilon^{ijk}.
\end{equation}
The matrix exponential can again be computed analytically, leading to
\begin{equation}
    \exp( i \rho^i T^i )_{kl} = \cos\qty(\rho) \delta^{kl} + \hat \rho^i  \varepsilon^{ikl} \sin\qty(\rho) + 2 \sin^2 \qty(\frac{\rho}{2}) \hat \rho^k \hat \rho^l,
\end{equation}
and
we see that in this case
\begin{equation}
    \gamma_A(\rho) = \int_{S^2}
    \frac{\dd{\Omega_\rho} }{4\pi}
    \frac{1}{3} \Tr\qty[   \exp( i \rho^i T^i ) ]
    = \frac{1}{3} \qty[1 + 2  \cos\qty( \rho)]
\end{equation}
and thus
\begin{equation}
\begin{split}
    \xi_A(\sigma) =& 4\pi \int_0^\infty \dd{\rho} \qty(\frac{\sigma \rho}{2\pi})^{3/2} J_{1/2}(\sigma \rho) \times 
     \frac{1}{3} \qty[1 + 2  \cos\qty( \rho)]\\
    =& - \frac{1}{3} \sigma \frac{\dd}{\dd{\sigma}} 
    \qty\Big[\delta\qty(\sigma) + 2\delta(\sigma-1)].
\end{split}
\end{equation}

\subsubsection{General representation}

Let us now compute $\gamma_R$ for a general representation.
We note that for the case $\nc =2$ the integral over the sphere $S^0= \qty{-1,1}$ in Eq.~\eqref{eq:gamma} reduces to summing over the points $-1$ and $1$, and the normalized eigenvalue vectors corresponding to these points are
$\hat \lambda(1) = \frac{1}{\sqrt{2}} (1,-1)$ and $\hat \lambda(-1) = \frac{1}{\sqrt{2}} (-1,1)$.
These are exactly the points on the sphere $S^0$ that are orthogonal to the vector $(1,1)$.
We then get
\begin{equation}
    \gamma_R( \rho )
    = \mathcal{N}
    \times\frac{1}{2}
    \sum_{p=\qty{-1,1}}
    \frac{1}{D_R}  \chi_R\qty[ \rho \sqrt{T_R} \hat \lambda(p) ]
      \times 2
      =  \mathcal{N}
    \sum_{p=\qty{-1,1}}
    \frac{1}{D_R}  \chi_R\qty[ \rho \sqrt{T_R} \hat \lambda(p) ].
\end{equation}
The normalization condition $\gamma_R(0)=1$ gives us $\mathcal{N}=\frac{1}{2}$.
It is now straightforward to verify the previous results for the fundamental and adjoint representations:
\begin{align}
    \gamma_F(\rho)
    =& \frac{1}{2} \qty[ \exp(  \frac{i\rho}{2} ) + \exp(-  \frac{i\rho}{2} ) ] 
    = \cos( \frac{\rho}{2} ) ,
    \\
    \gamma_A(\rho)
    =& \frac{1}{3} \qty[
    - 1
    + 2
    + 
    \exp(  i\rho ) + \exp(-  i\rho  ) ] 
    = \frac{1}{3}
    \qty[
    1
    + 2\cos( \rho )
    ].
\end{align}

\subsection{SU(3)}
\label{sec:su3}

For $\su(3)$ the evaluation of the matrix exponential is too involved, and thus we will only consider the method using representation theory.
The evaluate the function $\gamma_R$,
we first note that the integral in Eq.~\eqref{eq:gamma} is over the circle $S^1$.
The circle $S^1$ that is orthogonal to the vector $(1,1,1)$ can be parametrized
in terms of a single angle $\varphi = [0,2\pi)$
as
\begin{equation}
    \hat \lambda(\varphi) 
    = \sqrt{\frac{2}{3}} \qty(
    \cos \varphi, \cos \qty[\varphi + \frac{2\pi}{3}],  \cos \qty[\varphi - \frac{2\pi}{3}]
    ).
\end{equation}
To determine the normalization constant $\mathcal{N}$ in Eq.~\eqref{eq:gamma}, we first compute
\begin{equation}
        \prod_{i < j} \qty[ \hat \lambda_i(\varphi) -  \hat \lambda_j(\varphi)]^2 
        =\frac{1}{4} \qty[ 1-\cos(6 \varphi)]
\end{equation}
which has been simplified using basic trigonometric identities.
This leads to 
\begin{equation}
      \int_0^{2\pi}
    \frac{\dd{\varphi} }{2\pi}
        \prod_{i < j} \qty[ \hat \lambda_i(\varphi) -  \hat \lambda_j(\varphi)]^2
        = \frac{1}{4}
\end{equation}
and thus $\mathcal{N}=4$.
The expression for $\gamma_R$ can then be written as
\begin{equation}
    \gamma_R( \rho )
 =  \int_0^{2\pi}
      \frac{\dd{\varphi} }{2\pi}
      \times 
    \frac{1}{D_R}  \chi_R\qty[ \rho \sqrt{T_R} \hat \lambda(\varphi) ]
    \times 
    \qty[ 1-\cos(6 \varphi)].
\end{equation}

\subsubsection{Fundamental representation }

Substituting the character  for the fundamental representation from Eq.~\eqref{eq:character_F}, we end up with the integral
\begin{equation}
    \gamma_F( \rho )
 =  \int_0^{2\pi}
      \frac{\dd{\varphi} }{2\pi}
      \times 
  \frac{1}{3}
  \qty[
   e^{i
    \frac{\rho}{\sqrt{3}}
    \cos(\varphi)
    }
    +
    e^{i
    \frac{\rho}{\sqrt{3}}
    \cos(\varphi + 2\pi/3)
    }
    +
    e^{i
    \frac{\rho}{\sqrt{3}}
    \cos(\varphi - 2\pi/3)
    }
  ]
    \times 
    \qty[ 1-\cos(6 \varphi)].
\end{equation}
Using Eq.~\eqref{eq:expcossin} from Appendix~\ref{sec:angular_integral} this leads to
\begin{equation}
    \gamma_F(\sigma)
    =
     J_0\qty( \frac{\rho}{\sqrt{3}}) + J_6\qty( \frac{\rho}{\sqrt{3}}).
\end{equation}
The function $\xi_F$ can then be evaluated following Appendix~\ref{sec:bessel}:
\begin{equation}
\begin{split}
    \xi_F(\sigma)
    =&
       \vol{ S^{7} }
    \int_0^\infty \dd{\rho}  \qty(\frac{\sigma \rho}{2\pi})^{4}
    J_{3} \qty(\sigma \rho)
    \times
    \qty[
    J_0\qty( \frac{\rho}{\sqrt{3}}) + J_6\qty( \frac{\rho}{\sqrt{3}})
    ]\\
    =& \frac{1}{48} 
    \qty[ -\sigma^7 \qty(\frac{1}{\sigma} \dv{}{\sigma})^3 \frac{1}{\sigma} + \sigma \qty( \frac{1}{\sigma} \dv{}{\sigma} )^3 \sigma^5 ] 
   \delta\qty(\sigma - \frac{1}{\sqrt{3}})
\end{split}
\end{equation}
where the derivatives also act on the delta function.

\subsubsection{Adjoint representation  }

For the adjoint representation, after substituting the character~\eqref{eq:character_A} and simplifying a bit, we can write the result as the following integral
\begin{equation}
    \gamma_A( \rho )
 = \frac{1}{4} \int_0^{2\pi}
      \frac{\dd{\varphi} }{2\pi}
      \times 
 \Re  \qty[
 1 +
   e^{i
  \rho
    \sin(\varphi)
    }
    +
   e^{i
  \rho
    \sin(\varphi + 2\pi/3)
    }
    +
   e^{i
  \rho
    \sin(\varphi - 2\pi/3)
    }
  ]
    \times 
    \qty[ 1-\cos(6 \varphi)].
\end{equation}
Again, applying Eq.~\eqref{eq:expcossin} from Appendix~\ref{sec:angular_integral} we get
\begin{equation}
    \gamma_A( \rho )
 = \frac{1}{4}
 \qty[
 1 +
 3 J_0(\rho)
  - 3 J_6(\rho)
  ]
\end{equation}
and, following Appendix~\ref{sec:bessel},
\begin{equation}
    \begin{split}
          \xi_A(\sigma)
      =&
       \vol{ S^{7} }
    \int_0^\infty \dd{\rho}  \qty(\frac{\sigma \rho}{2\pi})^{4}
    J_{3} \qty(\sigma \rho)
    \times
  \frac{1}{4}
 \qty[
 1 +
 3 J_0(\rho)
  - 3 J_6(\rho)
  ]\\
    =&- \frac{1}{4 \times 48}
    \qty[\sigma^7 \qty(\frac{1}{\sigma} \dv{}{\sigma})^3 \frac{1}{\sigma} \qty\Big( \delta\qty(\sigma) +3\delta\qty(\sigma -1) ) + 3\sigma \qty( \frac{1}{\sigma} \dv{}{\sigma} )^3 \sigma^5 \delta\qty(\sigma - 1) ] .
    \end{split}
\end{equation}
The derivatives operate on everything to their right.

\subsection{\texorpdfstring{Large-$\nc$ limit
}{Large-Nc limit}
}

As the number of colors $\nc$ becomes large, computing the function $\gamma_R$ with Eq.~\eqref{eq:gamma} quickly becomes extremely tedious.
It is also very difficult to study the limit $\nc \to \infty$ using this expression.
For this reason, we consider a different method that allows us to compute an analytic expression for $\gamma_R$ in the limit $\nc \to \infty$.
It should also be noted that we will not present the functions $\xi_R$ due to the complexity of the integrals required to compute them.
Instead, we will only compute the function $\gamma_R$ that can be used to compute $\hat F^R$ for the sCGC model used in Sec.~\ref{sec:stable}.

To calculate the function
\begin{equation}
\label{eq:gamma2}
\gamma_R( \rho )
    =
    \int_{S^{D-1}} 
    \frac{\dd{\Omega_\rho}}{\vol{ S^{D-1} }}
    \frac{1}{D_R}
    \Tr[ \exp( i \rho^a t_R^a ) ]     
\end{equation}
in the large-$\nc$ limit, we will expand the exponential as a power series and compute each term separately.
This means that we need to compute the following series:
\begin{equation}
\label{eq:averaged_wilson_line_expand}
\gamma_R( \rho )
=   \int_{S^{D-1}} 
    \frac{\dd{\Omega_\rho}}{\vol{ S^{D-1} }}
    \frac{1}{D_R} \sum_{k=0}^\infty  \frac{i^k}{k!} \rho^{a_1} \ldots \rho^{a_k} \Tr \qty[ t_R^{a_1} \ldots t_R^{a_k}].
\end{equation}
It is possible to evaluate the integral over the angles of $\rho$ using
\begin{equation}
\label{eq:angle_integral}
\begin{split}
\int_{S^{D-1}} 
    \frac{\dd{\Omega_\rho}}{\vol{ S^{D-1} }}
 \rho^{a_1} \ldots \rho^{a_{2n}}
 =& 
 \rho^{2n} 
\times
 \frac{\Gamma \qty(\frac{D}{2}) \Gamma\qty(n+\frac{1}{2}) }{ \Gamma \qty(\frac{1}{2}) \Gamma \qty( \frac{D}{2} + n) }\\
 &\times \frac{1}{(2n-1)!!}  \sum_{\substack{\text{combinations } \qty{ b_{2k-1}, b_{2k} } = \qty{a_i,a_j} } } 
 \delta^{b_1 b_2} \ldots \delta^{b_{2n-1} b_{2n}} 
\end{split}
\end{equation}
for even powers of $\rho$.
In the sum, each pair $ \qty{ b_{2k-1}, b_{2k} }$ is considered only once:
for example, if $n=2$ the sum reduces to $\delta^{a_1 a_2} \delta^{a_3 a_4}+\delta^{a_1 a_3} \delta^{a_2 a_4}+\delta^{a_1 a_4} \delta^{a_2 a_3}$.
Equation~\eqref{eq:angle_integral} can be derived by noting that the only possible color indices are the Kronecker deltas, and the overall normalization can be recovered by considering some special case such as $a_1 = a_2 = \ldots = a_{2n}$.
For odd powers of $\rho$, the angular integral vanishes by symmetry.

The color structure of Eq.~\eqref{eq:angle_integral} means that we need to consider every contraction of the matrices $t^a_R$ in the trace.
Let us denote this by $\sym(n)$ where $n$ is the number of contractions.
For example, for $n=2$ this is
\begin{equation}
    \sym_R(2) = \Tr \big[  
     \wick{\c1 t \, \c1 t \, \c1 t \, \c1 t }  
  \big]
  +
  \Tr \big[  
     \wick{\c2 t \, \c1 t \, \c1 t \, \c2 t }
  \big]
  +
  \Tr \big[  
     \wick{\c1 t \, \c2 t \, \c1 t \, \c2 t }
  \big].
\end{equation}
We can then simplify Eq.~\eqref{eq:averaged_wilson_line_expand} to
\begin{equation}
\label{eq:gamma_sym}
\gamma_R( \rho )
= 
    \frac{1}{D_R} \sum_{n=0}^\infty  \frac{(-1)^n}{(2n)!}  \rho^{2n} 
\times
 \frac{\Gamma \qty(\frac{D}{2}) \Gamma\qty(n+\frac{1}{2}) }{ \Gamma \qty(\frac{1}{2}) \Gamma \qty( \frac{D}{2} + n) }
 \times \frac{ \sym_R(n)}{(2n-1)!!} 
 .
\end{equation}
Note that so far we have not made any approximations.

\subsubsection{Fundamental representation}
\label{app:fundamental_largeNc}

We still need to compute $\sym_R(n)$. 
For the fundamental representation, we can apply the Fierz identity~\eqref{eq:fierz}
to compute $\sym_F(n)$ for any finite $n$ directly. 
However, as $n$ grows the number of terms in $\sym_F(n)$ grows exponentially, and this method becomes infeasible.
Fortunately, things simplify in the large-$\nc$ limit.

First, we note that we can drop the second term in the Fierz identity~\eqref{eq:fierz} as it is suppressed by $1/\nc$.
Second, we can also drop any term in $\sym_F(n)$ where the contractions of the color matrices cross.
The reasoning for this is that we want every contraction to split the trace into two different pieces, as each trace corresponds to a factor of $\nc$.
On the other hand, crossing contractions means that at some point we have to contract matrices in separate pieces; this would then recombine those traces, dropping one factor of $\nc$.
An illustration of this is the following term:
\begin{equation}
      \Tr \big[\wick{\c1 t \, \c2 t \, \c3 t \, \c1 t \, \c2 t \, \c3 t } \big]
      \approx \frac{1}{2} 
      \Tr \big[\wick{\c1 t \, \c2 t  \big]
      \Tr \big[\c1 t \, \c2 t } \big]
       \approx \frac{1}{4} 
      \Tr \big[\wick{\c1 t \, \c1 t } \big]
       \approx \frac{\nc^2}{8}
\end{equation}
which can be compared to
\begin{equation}
      \Tr \big[\wick{\c2 t \, \c1 t \, \c1 t \, \c2 t \, \c1 t \, \c1 t } \big]
       \approx \frac{1}{2} 
      \Tr \big[\wick{\c1 t\, \c1 t } \big]
       \Tr \big[\wick{\c1 t  \, \c1 t } \big]
       \approx \frac{\nc^4}{8}
\end{equation}
where the approximation signs correspond to keeping only the leading terms in $\nc$.
Dropping the terms with crossing contractions simplifies the situation as all of the remaining terms yield the same result:
\begin{equation}
     \Tr \big[\wick{\c1 t \, \c1 t \ldots \c1 t \, \c1 t } \big]
     \approx \frac{\nc^{n+1}}{2^n}
\end{equation}
where $n$ is the number of contractions.
This results from the fact that without crossings, we must have some term in the trace where we contract neighboring color matrices.
Using the Fierz identity, we can replace this with
\begin{equation}
    \ldots t^a \wick{\c1 t\, \c1 t } t^b \ldots \approx \frac{\nc}{2} \ldots t^a  t^b \ldots
\end{equation}
and as the resulting trace cannot have any crossings either, we can continue this until we have no contractions left.
The final power of $\nc$ comes from the trace itself.
We can then write
\begin{equation}
    \sym_F(n) \approx C(n) \frac{\nc^{n+1}}{2^n}
\end{equation}
where $C(n)$ is the  number of terms with no crossing contractions.

To compute the number $C(n)$, we will write it in terms of a recursion relation that can be solved.
To construct this relation, we start by considering all possible contractions for the first color matrix $t^a$ in a trace.
This first matrix can be contracted with any other matrix provided that there is an even number of matrices between the two, as otherwise there would have to be crossing contractions.
Contracting the first matrix then  divides the trace into two separate traces, for which we can again count the number of possible non-crossing contractions.
These separate traces cannot have contractions connecting them: otherwise, there would be crossing contractions in the initial trace which is not allowed.
Thus, we can  count the number of possible contractions in these two traces separately.
This leads to the recursion relation for the number $C(n)$:
\begin{equation}
   \label{eq:catalan_recursion}
    C(n) = \sum_{k=0}^{n-1} C(k) C(n-1-k)
\end{equation}
where each term in the sum corresponds to having $2k$ matrices between the first matrix and its contracted counterpart.
The initial condition for this recursion relation is $C(0)=1$, which corresponds to an ``empty'' trace.
The recursion relation~\eqref{eq:catalan_recursion} is well known and is solved by the Catalan numbers
\begin{equation}
    C(n) = \frac{(2n)!}{n!(n+1)!}.
\end{equation}

At this point, we can write the expression for $\gamma_F$ as
\begin{equation}
\label{eq:gamma_sym_F}
\gamma_F( \rho )
\approx 
    \frac{1}{\nc} \sum_{n=0}^\infty  \frac{(-1)^n}{(2n)!}  \rho^{2n} 
\times
 \frac{\Gamma \qty(\frac{D}{2}) \Gamma\qty(n+\frac{1}{2}) }{ \Gamma \qty(\frac{1}{2}) \Gamma \qty( \frac{D}{2} + n) }
 \times \frac{ 1}{(2n-1)!!}  \times C(n) \frac{\nc^{n+1}}{2^n}
 .
\end{equation}
When employing the large-$\nc$ limit for each trace, we should also approximate $\nc \gg n$ for each term in the series. 
We then get
\begin{equation}
\label{eq:large_Nc_gamma_function}
    \frac{\Gamma \qty( \frac{D}{2})}{ \Gamma \qty( \frac{D}{2} + n)} \approx \qty(\frac{D}{2})^{-n} \approx \qty(\frac{\nc^2}{2})^{-n} 
\end{equation}
which can be verified by noting that $D = \nc^2- 1 \approx \nc^2$ and using Stirling's approximation for the gamma function.
This leads to
\begin{equation}
\label{eq:gammaF_largeNc}
\begin{split}
   \gamma_F(\rho) \approx& \sum_{n=0}^\infty \frac{(-1)^n}{(2n)!} \times \rho^{2n} \times  \frac{\Gamma \qty(n+ \frac{1}{2})}{\Gamma \qty(\frac{1}{2})} \qty(\frac{2}{\nc^2})^n
    \times \frac{1}{(2n-1)!!} \times  C(n)  \qty(\frac{\nc}{2})^n \\
    =&  \sum_{n=0}^\infty \frac{C(n)}{(2n)!} \qty(\frac{-\rho^2}{2\nc})^{n}
    =  \sum_{n=0}^\infty \frac{1}{n! (n+1)!} \qty(\frac{-\rho^2}{2\nc})^{n}
    = \frac{\sqrt{2 \nc}}{\rho} J_1 \qty( \rho \sqrt{\frac{2}{\nc} })
\end{split}
\end{equation}
where we have simplified the terms in the series by using properties of the gamma function.
The approximations correspond to equalities in the limit $\nc \to \infty$.

\subsubsection{Adjoint representation}

In the adjoint representation, the color matrices can be written as
\begin{equation}
    \qty[T^a]_{bc}= -2 \Tr( t^a [t^b, t^c] ).
\end{equation}
Before the angular integration, we have terms like
\begin{equation}
\label{eq:chi_n}
\begin{split}
\chi_{n}(\vec \rho)=&
   \rho^{a_1} \ldots \rho^{a_n} \Tr( T^{a_1} \ldots  T^{a_n} ) \\
   =& (-2)^n   \rho^{a_1} \ldots \rho^{a_n}   
   \Tr( t^{a_1} [t^{b_n}, t^{b_1}] )  \Tr( t^{a_2} [t^{b_1}, t^{b_2}] ) \ldots  \Tr( t^{a_n} [t^{b_{n-1}}, t^{b_n}] )
\end{split}
\end{equation}
in the series in Eq.~\eqref{eq:averaged_wilson_line_expand}.
Let us denote $\rho^a t^a = \slashed{\rho}$ to simplify expressions.
Using the Fierz identity~\eqref{eq:fierz} for $t^{b_i} \otimes t^{b_i}$ we can combine the traces in Eq.~\eqref{eq:chi_n} together.
For the first few $n$ we can then simplify $\chi_n$ to:
\begin{align}
    \chi_1(\vec \rho) &= 2\times (-1) \times\delta^{b_1 b_n} \Tr( t^{b_n}  \qty[t^{b_1} \slashed{\rho} - \slashed{\rho} t^{b_1} ]  )\\
    \chi_2(\vec \rho) &= 2\times (-1)^2\times \delta^{b_2 b_n}\Tr( t^{b_n}  
    \qty[t^{b_2} \slashed{\rho} \slashed{\rho} - 2 \slashed{\rho} t^{b_2} \slashed{\rho} + \slashed{\rho}\slashed{\rho}] t^{b_2}   )\\
    \chi_3(\vec \rho) &= 2\times (-1)^3 \times\delta^{b_3 b_n}\Tr( t^{b_n}  
    \qty[t^{b_3} \slashed{\rho} \slashed{\rho} \slashed{\rho}
    -3 \slashed{\rho} t^{b_3}\slashed{\rho} \slashed{\rho}
    +3\slashed{\rho} \slashed{\rho}t^{b_3} \slashed{\rho}
    - \slashed{\rho} \slashed{\rho} \slashed{\rho}t^{b_3}
    ]  )
\end{align}
where we have not done the final contraction to illustrate the general pattern:
up to a minus sign, the coefficients for the terms correspond to the binomial coefficients.
This can be easily understood by noting that each time we use the Fierz identity for a contraction, we essentially replace
\begin{equation}
    t^b \to \frac{1}{2} \qty[ t^b \slashed{\rho} - \slashed{\rho} t^b ].
\end{equation}
Thus, we can generalize this and write
\begin{equation}
    \chi_n(\vec \rho) =2\times  (-1)^n \times \Tr( t^b \qty[ \sum_{k=0}^n (-1)^k \binom{n}{k} \slashed{\rho}^k t^b \slashed{\rho}^{n-k}  ] ).
\end{equation}
Note that so far we have not made any assumptions $\nc$ and thus this result is completely general.
However, to proceed further it is convenient to do the angular integral~\eqref{eq:angle_integral} and invoke the large-$\nc$ limit.

First of all, we note that in the large-$\nc$ limit we can write
\begin{equation}
    \chi_n(\vec \rho) \approx  \sum_{k=0}^n (-1)^{n+k}\binom{n}{k} \Tr(  \slashed{\rho}^k ) \Tr( \slashed{\rho}^{n-k}   ).
\end{equation}
We follow the steps in Section~\ref{app:fundamental_largeNc} and note that we only need to keep the contractions that separate the trace further.
This yields
\begin{equation}
\label{eq:adjoint_fn}
\begin{split}
    \overline \chi_{2n}(\rho) =& 
     \int_{S^{D-1}}  \frac{\dd{\Omega_\rho}}{\vol{S^{D-1}}} \chi_{2n}(\vec \rho) \\
     \approx&  \rho^{2n} 
\times
 \frac{\Gamma \qty(\frac{D}{2}) \Gamma\qty(n+\frac{1}{2}) }{ \Gamma \qty(\frac{1}{2}) \Gamma \qty( \frac{D}{2} + n) }\times \frac{1}{(2n-1)!!} 
 \times
 \sum_{l=0}^{n}  \binom{2n}{2l} \, \sym_F\qty(   l )  \, \sym_F\qty( n-l )\\
     \approx&  
     \rho^{2n} 
\times
 \frac{\Gamma \qty(\frac{D}{2}) \Gamma\qty(n+\frac{1}{2}) }{ \Gamma \qty(\frac{1}{2}) \Gamma \qty( \frac{D}{2} + n) }\times \frac{1}{(2n-1)!!} 
 \times
     \frac{\nc^{n+2}}{2^{n}} \sum_{l=0}^{n}  \binom{2n}{2l} \, C\qty( l )  \, C\qty( n-l )\\
     \approx&  
   \nc^2
     \qty(\frac{\rho^2}{2 \nc})^n
        \times 
     \sum_{l=0}^{n}  \binom{2n}{2l} \, C\qty( l )  \, C\qty( n-l )
\end{split}
\end{equation}
At this point, we note that we can write $\gamma_A$ in the form
\begin{equation}
\gamma_A( \rho )
\approx   \frac{1}{\nc^2} \sum_{n=0}^\infty  \frac{(-1)^n}{(2n)!} 
\overline \chi_{2n}(\rho)
\approx  \sum_{n=0}^\infty  \frac{1}{(2n)!} 
\qty(-\frac{\rho^2}{2 \nc})^n
        \times 
     \sum_{l=0}^{n}  \binom{2n}{2l} \, C\qty( l )  \, C\qty( n-l ).
\end{equation}
The coefficients in the final line are reminiscent of the ones for the series of $\gamma_F$ in Eq.~\eqref{eq:gammaF_largeNc}.
It turns out that we can write this result in terms of $\gamma_F$ by using the Cauchy product form for the Taylor series:
\begin{equation}
\label{eq:cauchy_product}
     \sum_{k=0}^\infty \frac{1}{(2k)!} a_k x^{k} \times \sum_{l=0}^\infty \frac{1}{(2l)!} b_l x^{l}
    = \sum_{n=0}^\infty \frac{1}{(2n)!} x^{n} \sum_{l=0}^n \binom{2n}{2l} a_l b_{n-l}.
\end{equation}
Using this for $\gamma_A$ directly gives us
\begin{equation}
\gamma_A( \rho )
\approx
\qty[
\sum_{k=0}^\infty \frac{1}{(2k)!} C(k) \qty(- \frac{\rho^2}{2\nc})^{k}
]^2
\approx
\gamma_F^2(\rho)
\approx \frac{2 \nc}{\rho^2 } J_1\qty(\rho \sqrt{\frac{2}{\nc}})^2,
\end{equation}
where again the approximations are exactly valid in the limit $\nc \to \infty$.

\section{Useful integrals}
\label{app:integrals}

\subsection{Bessel function integrals}
\label{sec:bessel}

When computing the function $\gamma_R$ for various representations, we often encounter integrals of the following type:
\begin{equation}
\label{eq:bessel_integral}
    \int_0^\infty \dd{\rho} \rho^{1+k}
    J_{\alpha}(a \rho)
    J_{\alpha + n}(b \rho),
\end{equation}
where $a, b \geq 0$, and $k$ and $n$ are integers that have the same parity and satisfy $n \geq \abs{k}$.
These types of integrals can be computed using the following identities:
\begin{align}
\label{eq:bessel_orthogonality}
    \int_0^\infty \dd{\rho} \rho  J_{\alpha}(a \rho)
    J_{\alpha}(b \rho) &= \frac{1}{a} \delta(a -b), \\ 
\label{eq:bessel_up}
    \frac{1}{c} \partial_c \qty[ c^{-\nu} J_\nu(c \rho)  ] 
    &= - c^{- (\nu + 1)} \rho J_{\nu + 1}(c \rho)  ,
 \\
\label{eq:bessel_down}
    \frac{1}{c} \partial_c \qty[ c^\nu J_\nu(c \rho)  ] 
    &=  c^{\nu -1} \rho J_{\nu -1}(c \rho).
\end{align}
We wish to transform the integral~\eqref{eq:bessel_integral} into a form where we can use the orthogonality relation~\eqref{eq:bessel_orthogonality},
which can be achieved by raising or lowering the order of the Bessel functions using Eqs.~\eqref{eq:bessel_up} and~\eqref{eq:bessel_down}.
As can be verified by a direct computation, this leads to the equation
\begin{equation}
\label{eq:bessel_integral_evaluated}
    \begin{split}
          &\int_0^\infty \dd{\rho} \rho^{1+k}
    J_{\alpha}(a \rho)
    J_{\alpha + n}(b \rho) \\
    =&
    a^{-\alpha} b^{\alpha+n}
    \qty[ \frac{1}{a} \partial_a]^{(n+k)/2}
    \qty[- \frac{1}{b} \partial_b]^{(n-k)/2}
    \int_0^\infty \dd{\rho} \rho
    \qty(\frac{a}{b})^{\alpha + (n+k)/2}
    J_{\alpha + (n+k)/2}(a \rho)
    J_{\alpha + (n+k)/2}(b \rho) \\
    =&
    a^{-\alpha} b^{\alpha+n}
    \qty[ \frac{1}{a} \partial_a]^{(n+k)/2}
    \qty[- \frac{1}{b} \partial_b]^{(n-k)/2}
\qty{
    \qty(\frac{a}{b})^{\alpha + (n+k)/2}
    \times
\frac{1}{a} \delta(a-b)
} \\
    =&
    a^{-\alpha} 
    \qty[ \frac{1}{a} \partial_a]^{n}
\qty{
a^{\alpha+n-1} \delta(a-b)
}
    =
    b^{\alpha+n} 
    \qty[- \frac{1}{b} \partial_b]^{n}
\qty{
b^{-\alpha-1} \delta(a-b)
}.
    \end{split}
\end{equation}
Going to the last line can be proven by integrating this expression over $a$ and $b$ with a test function and integrating by parts.
Partial integration in terms of $b$ leads to the first form in the last line, whereas partial integration in terms of $a$ yields the second form.

\subsection{Angular integrals}
\label{sec:angular_integral}

The following integrals turn out to be useful:
\begin{align}
\label{eq:expcossin}
  \int_0^{2\pi} \dd{\varphi} e^{i z \cos\qty(\varphi-\delta)}  \cos(n \varphi)
  = \int_0^{2\pi} \dd{\varphi} e^{i z \sin\qty(\varphi-\delta + \pi/2)}  \cos(n \varphi)
    = 2\pi i^n \cos(n \delta) J_n(z)  
\end{align}
where $n$ is an integer, and $z$ and $\delta$ are real numbers.
These integrals can be proven using the identity (see Eq.~(3.915-2) in Ref.~\cite{MR2360010})
\begin{align}
  \int_0^{\pi} \dd{\varphi} e^{i z \cos \varphi}  \cos(n \varphi)
    = \pi i^n  J_n(z)  ,
\end{align}
and noting that
\begin{equation}
\begin{split}
      &\int_0^{2\pi} \dd{\varphi} e^{i z \cos\qty(\varphi-\delta)}  \cos(n \varphi)
  = \int_{-\delta}^{2\pi-\delta} \dd{\varphi} e^{i z \cos \varphi} \cos(n \qty[\varphi + \delta])\\
  =& \int_{-\pi}^{\pi} \dd{\varphi} e^{i z \cos \varphi} \cos(n \qty[\varphi + \delta])
  = \int_{-\pi}^{\pi} \dd{\varphi} e^{i z \cos \varphi}
  \qty[\cos(n \varphi) \cos( n \delta) - \sin(n \varphi) \sin( n \delta)]\\
  =& \int_{-\pi}^{\pi} \dd{\varphi} e^{i z \cos \varphi}
  \cos(n \varphi) \cos( n \delta)
    = 2 \cos( n \delta)   \int_{0}^{\pi} \dd{\varphi} e^{i z \cos \varphi}
  \cos(n \varphi).
\end{split}    
\end{equation}

\section{\texorpdfstring{Computing $\hat F^R$ for the stable color-glass condensate model}
{
Computing F for the stable color-glass condensate model}}
\label{app:Fhat}

When considering the sCGC model with $\w_z(\sigma^2)= \abs{ \mu(z) \sigma}^\alpha$,
the dependence on the representation appears as a constant that can be expressed as an integral over $\gamma_R$.
That is, we wish to compute
\begin{equation}
\label{eq:Fz_stable}
\begin{split}
    F^R_z(\xt ,\yt)
    =& \int_0^\infty \dd{\sigma} \xi_R(\sigma)
    \abs{ \frac{\gs}{2\pi} K_{\xt \yt \zt} \mu(z) \sigma}^\alpha \\
    =&  \abs{ \frac{\gs}{2\pi} K_{\xt \yt \zt} \mu(z)}^\alpha
    \times 
       \vol{S^{D-1}} 
    \int_0^\infty \dd{\sigma} \dd{\rho} 
    \qty( \frac{\sigma \rho}{2\pi} )^{\frac{D}{2}}
    \sigma^\alpha
    J_{D/2-1}(\sigma \rho) 
    \gamma_R(\rho).
\end{split}
\end{equation}
The final integrals over $\rho$ and $\sigma$ can be computed by noting that they have a connection to the $D$-dimensional fractional Laplace operator.
The fractional Laplacian of a function $g$ can be written in terms of its Fourier transform $\tilde g$ as~\cite{MR3613319}
\begin{equation}
   \abs{D}^\alpha g(\vec x)
    \equiv 
   - \int \frac{\dd[D]{\vec k}}{(2\pi)^D} e^{i \vec x \vdot \vec k}
     k^\alpha \tilde g(\vec k)
     =-
     \int \frac{\dd[D]{\vec k} \dd[D]{\vec y}}{(2\pi)^D}
     e^{i  (\vec x -\vec y) \vdot \vec k}
     k^\alpha g(\vec y),
\end{equation}
where $k^2 = \vec k^{\, 2}$.
If we now assume that $g$ only depends on the norm $x^2$, the fractional Laplacian at $\vec x = 0$ becomes
\begin{equation}
    \abs{D}^\alpha g(0)
     =-
     \vol{S^{D-1}} 
     \int_0^\infty \dd{k} \dd{y}
     \qty(\frac{k y}{2\pi})^{D/2} 
    k^\alpha
     J_{D/2 - 1}(k y)
      g(y).
\end{equation}
This integral has the same form as Eq.~\eqref{eq:Fz_stable}, and thus
\begin{equation}
    F^R_z(\xt,\yt)
    =  \abs{ \frac{\gs}{2\pi} K_{\xt \yt \zt} \mu(z)}^\alpha
    \hat F_R(\alpha)
\end{equation}
where
\begin{equation}
\label{eq:Fhat}
    \hat F_R(\alpha)
    = \int_0^\infty \dd{\sigma} \sigma^\alpha \xi_R(\sigma)
    = - \abs{D}^\alpha \gamma_R(0).
\end{equation}

\subsection{Evaluating the fractional Laplacian}
\label{app:riesz}

Let us now show how to relate the fractional Laplacian to the coefficients of the Taylor series of the function $g$ under some quite general assumptions.
That is, we will assume that the function $g$ is analytic and has the Taylor expansion
\begin{equation}
\label{eq:g_taylor}
    g(x) = \sum_{n=0}^\infty c_n \qty(-x^2)^n
\end{equation}
where the coefficients $c_n$ can be written in terms of a function $c(z)$ such that $c(z) = c_z$ when $z \in \mathbb{N}$.
At this point, we  note that the fractional Laplacian can be written in terms of the Mellin transform:
\begin{equation}
\begin{split}
   \abs{D}^\alpha g(0)
  &=-
   \vol{S^{D-1}} 
     \int_0^\infty \frac{\dd{y}}{y}
     \frac{ 1}{\pi^{D/2}}
       \qty(\frac{2}{y})^{\alpha}
    \frac{\Gamma\qty( \frac{D+\alpha}{2})}{\Gamma\qty(-\frac{1}{2}\alpha)} 
      g(y) \\
  &=-
  \vol{S^{D-1}} 
  \frac{2^\alpha}{\pi^{D/2}} \frac{\Gamma\qty( \frac{D+\alpha}{2})}{\Gamma\qty(-\frac{1}{2}\alpha)} \hat g(-\alpha),    
\end{split}
\end{equation}
where we have used the distributional identity 
\begin{equation}
    \int_0^\infty \dd{k} k^{D/2+\alpha} J_{D/2-1}(ky) =  \frac{1}{y} \qty(\frac{2}{y})^{\alpha+D/2}
    \frac{\Gamma\qty( \frac{D+\alpha}{2})}{\Gamma\qty(-\frac{1}{2}\alpha)} 
\end{equation}
and the Mellin transform is defined as
\begin{equation}
    \hat g(s) = \int_0^\infty \frac{ \dd{x}}{x} x^{s} g(x).
\end{equation}

\begin{figure}
    \centering
    \begin{overpic}[width=0.5\textwidth]{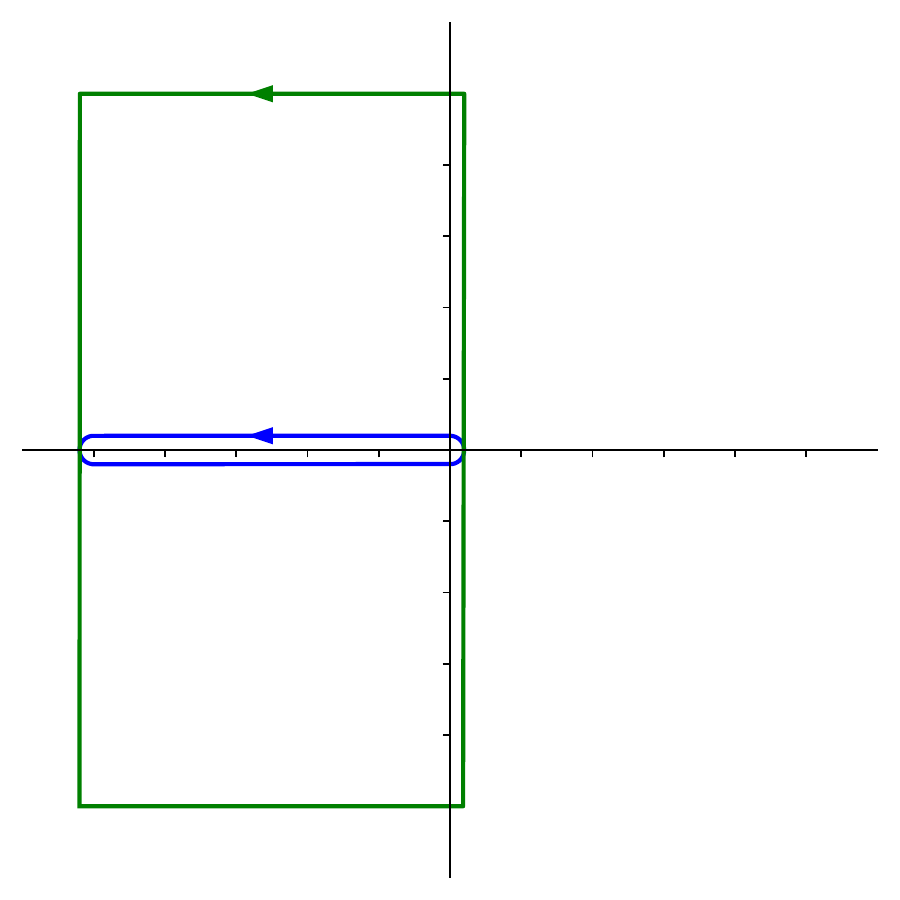}
        \put (25,93) {\color{Green4} $C_n'$}
        \put (25,55) {\color{blue} $C_n$}
        \put (8,44) {$-n$}
        \put (47,44) {$0$}
        \put (100,44) {$\Re z$}
        \put (47,100) {$\Im z$}
    \end{overpic}
    \caption{Contours for evaluating the fractional Laplacian.
    }
    \label{fig:riesz_contours}
\end{figure}
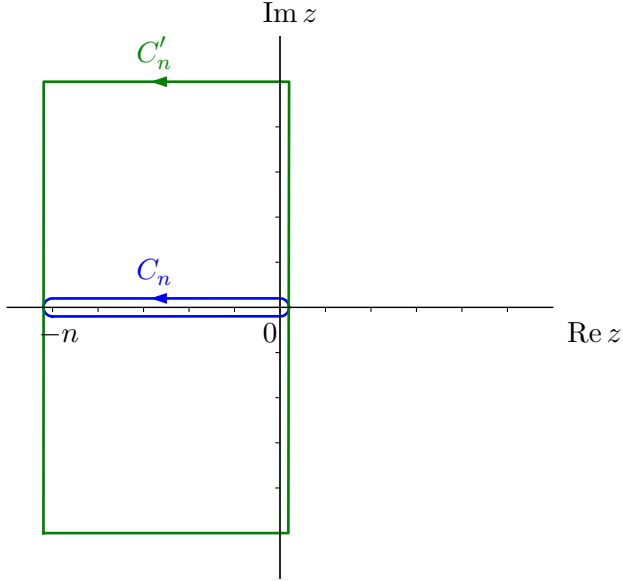

We now want to figure out the Mellin transform of $g$ using the Taylor expansion~\eqref{eq:g_taylor}.
To this end, we will rewrite it as
\begin{equation}
\label{eq:g_integral}
    g(x) = \lim_{n \to \infty} \int_{C_n} \frac{\dd{z}}{2\pi i} \frac{\pi}{2\sin(\pi z/2)} c\qty(-\frac{z}{2}) x^{-z},
\end{equation}
where the contour $C_n$ circles the negative real axis counter-clockwise as shown in Fig.~\ref{fig:riesz_contours}. 
This rewriting relies on the assumption that $c(z)$ is analytic around the positive real axis.
The Mellin transform is then given by
\begin{equation}
    \hat g(s)
    = \lim_{n \to \infty} \int_0^\infty \dd{x}  \int_{C_n} \frac{\dd{z}}{2\pi i}  \frac{\pi}{2\sin(\pi z/2)} c\qty(-\frac{z}{2}) x^{-z+s-1}.
\end{equation}
We will now denote 
\begin{equation}
    \hat h(z)
    =  \frac{\pi}{2\sin(\pi z/2)} c\qty(-\frac{z}{2}).
\end{equation}
Assuming that $c(z)$ is analytic in the whole positive real half-plane (such that $c(-z/2)$ is analytic in the negative real half-plane), we can deform the contour $C_n$ to a new contour $C'_n$ where we push the horizontal lines to infinity as illustrated in Fig.~\ref{fig:riesz_contours}.
Assuming also that $c(-z/2)$ increases slowly enough as $\Im z \to \pm \infty$ or $\Re z \to -\infty$, we can ignore most of the contour $C'_n$ and focus only on the integral along the imaginary axis. 
We then get
\begin{equation}
    \hat g(s)
    = \int_0^\infty \dd{x}  \int_{-i\infty}^{+i\infty} \frac{\dd{z}}{2\pi i}  x^{-z+s-1} \hat h(z),
\end{equation}
and we can note that the first integral corresponds to an inverse Mellin transform
\begin{equation}
    h(x) = \int_{-i\infty}^{+i\infty}  \frac{\dd{z}}{2\pi i} x^{-z} \hat h(z),
\end{equation}
allowing us to write
\begin{equation}
    \hat g(s)
    = \int_0^\infty \dd{x}  x^{s-1}  h(x).
\end{equation}
This is just the Mellin transform of $h(x)$, i.e. 
\begin{equation}
    \hat g(s)
    =  \hat h(s) 
    = \frac{\pi}{2\sin(\pi z/2)} c\qty(-\frac{z}{2}).
\end{equation}
Our final expression for the fractional Laplacian is then
\begin{equation}
\label{eq:riesz_relation}
   \abs{D}^\alpha g(0)
  = -
  \vol{S^{D-1}} 
  \frac{2^\alpha}{\pi^{D/2}} \frac{\Gamma\qty( \frac{D+\alpha}{2})}{\Gamma\qty(-\frac{1}{2}\alpha)}  \frac{\pi}{2\sin(-\pi \alpha/2)} c\qty(\frac{\alpha}{2})
  = - 2^{\alpha} \frac{\Gamma\qty(\frac{D+\alpha}{2}) \Gamma\qty(1 +\frac{\alpha}{2})}{\Gamma\qty(\frac{D}{2})}
    c \qty(\frac{\alpha}{2}),
\end{equation}
where in the last step we used the inversion formula for the gamma function, along with the volume of the $(D-1)$-sphere
\begin{equation}
    \vol{S^{D-1}} = \frac{2 \pi^{D/2}}{\Gamma\qty(\frac{D}{2})}.
\end{equation}
Equation~\eqref{eq:riesz_relation} can then be used to compute $\hat F$ given that we know the Taylor expansion of $\gamma_R$.

Using this equation, we should be careful to make sure that the assumptions in its derivation are valid.
That is, we need to assume that the function $c(z)$ corresponding to the coefficients in the Taylor expansion satisfies the following criteria:
\begin{enumerate}
    \item The function $c(z)$ is analytic in the region $\Re z \geq 0$.
    \item \label{it:bounded} The function $c(z)/\sin(\pi z)$ decreases fast enough when $\abs{z} \to \infty$ in the positive real half plane, such that we can ignore the contours at the infinity.
\end{enumerate}
These criteria are satisfied for the functions considered in this work.

\subsection{Results for various representations}

\begin{table}[t]
  \renewcommand{\arraystretch}{2}
    \centering\begin{tabular}{ccc}
\toprule 
 Representation & $c_R(z)$   & $\hat F_R(\alpha)$
\\ \midrule
$\su_F(2)$  & 
$\frac{1}{4^z}\times \frac{1}{\Gamma( 2z+1 )} $  & 
$\frac{1}{2^\alpha} \times  (1+\alpha)$ 
\\
$\su_A(2)$   &
$ \frac{2}{3} \times  \frac{1}{\Gamma( 2z+1 )} $    & 
$  \frac{2}{3} \times (1+\alpha)$
\\
$\su_F(3)$     & 
$ \frac{1}{12^z} \times \qty[ \frac{1}{\Gamma(z+1)^2} - \frac{1}{\Gamma(z-2) \Gamma( z+4 )} ] $&
$ \frac{1}{6 \times 3^{\alpha/2}}
\times
\qty[
    \frac{\Gamma\qty(4+\frac{\alpha}{2})}{\Gamma\qty(1+\frac{\alpha}{2})}
    -
    \frac{\Gamma\qty(1+\frac{\alpha}{2})}{\Gamma\qty(-2+\frac{\alpha}{2})}
    ]$ \\
$\su_A(3)$     & 
$\frac{3}{4 \times 4^z} \times \qty[ \frac{1}{\Gamma(z+1)^2} + \frac{1}{\Gamma(z-2) \Gamma( z+4 ) } ] $&
$  \frac{1}{8}
\times
    \qty[
    \frac{\Gamma\qty(4+\frac{\alpha}{2})}{\Gamma\qty(1+\frac{\alpha}{2})}
    +
    \frac{\Gamma\qty(1+\frac{\alpha}{2})}{\Gamma\qty(-2+\frac{\alpha}{2})}
    ]$ \\
$\su_F(\infty)$     & 
$
\frac{1}{(2\nc)^z} \times
\frac{1}{\Gamma(z+1) \Gamma(z+2)}  $&
$\frac{1}{\Gamma\qty(2 + \frac{\alpha}{2})} \times \nc^{\alpha/2}$ \\
$\su_A(\infty)$     & 
$
\frac{1}{(2\nc)^z} \times
\frac{\Gamma\qty(2z + 3)}{\Gamma(z+1) \Gamma(z+2)^2  \Gamma(z+3)  }
$&
$   \frac{\Gamma\qty(3 +\alpha)}{\Gamma\qty(2 + \frac{\alpha}{2})^2\Gamma\qty(3 + \frac{\alpha}{2})} 
 \times    \nc^{\alpha/2}$ \\
\bottomrule
\end{tabular}
    \caption{The coefficient function $c_R(z)$ and the function $\hat F_R(\alpha)$ for different groups $\su(\nc)$ and their representations.
    Note that for the adjoint representations $\su_A(2)$ and $\su_A(3)$ we drop the constant term when defining $c_R(z)$.
    }
    \label{tab:F_hat}
\end{table}

The results for $\gamma_R$ computed in Appendix~\ref{app:moment-generating_function} now allow us to compute the function $\hat F_R$ using the fractional Laplacian.
The results are  collected in Table~\ref{tab:F_hat} where we also show the suitable function $c_R(z)$ that defines the Taylor series expansion for $\gamma_R$:
\begin{equation}
\label{eq:gamma_taylor}
    \gamma_R(\rho) = c'_R + \sum_{n=0}^\infty c_R(n) \qty(-\rho^2)^n.
\end{equation}
We note that here we also include an additional constant $c_R'$:
this is needed for the adjoint representations $\su_A(2)$
and $\su_A(3)$ for which this constant is $1/3$ and $1/4$, respectively.
While one could try to include such a term in the function $c_R(z)$ by adding a term $c_R' /\Gamma(-z+1)$, this leads to an incorrect result:
such a function violates the criterion~\ref{it:bounded} for $c_R(z)$
because the function 
$1/\qty[\Gamma(-z+1) \sin(\pi z)] = \pi \Gamma(z)$
does not decrease fast enough for $z \to \infty$, invalidating the final result.
Thus, it is easier to isolate this constant term $c'_R$ from $\gamma_R$ and 
compute the integral~\eqref{eq:Fz_stable} for it by other means.
As can be explicitly verified, the $\rho$ integral for this term corresponds to a delta function $\delta(\sigma)$ such that we are left with the integral
\begin{equation}
    \int_0^\infty \dd{\sigma} \sigma^\alpha \times c_R' \delta(\sigma)
    = 0
\end{equation}
which vanishes for the allowable values $0 < \alpha \leq 2$.
We can then ignore the constant term $c'_R$ so that all of the relevant information is encoded in $c_R$.

Finally, we note some special values of $\hat F_R$ that work as a crosscheck of the results in Table~\ref{tab:F_hat}:
\begin{align}
    \lim_{\alpha \to 0} \hat F_R(\alpha) &= 
    \begin{cases}
        1 &\text{ for $R =F$} \\
       1 - \frac{\nc-1}{ \nc^2 -1} = \frac{\nc}{\nc+1} &\text{ for $R =A$} \\
    \end{cases} 
    ,
    \\
    \hat F_R(2) &= C_R.
\end{align}
The second identity, for $ \hat F_R(2)$, is the Gaussian case: 
one way to derive this result is to note that for $\alpha=2$ the fractional Laplacian in Eq.~\eqref{eq:Fhat} corresponds to a second derivative which can be computed from e.g. Eq.~\eqref{eq:gamma}.
The first case, $    \lim_{\alpha \to 0} \hat F_R(\alpha)$, is a little bit trickier:
if we were to compute $\hat F_R(0)$ directly using Eq.~\eqref{eq:gamma_normlization}, we would get $\hat F_R(0) =1$ instead.
The reason why this differs from the limit $\alpha \to 0$ is because of the extra term $c'_R$ in Eq.~\eqref{eq:gamma_taylor} for the adjoint representation that only contributes for $\alpha=0$.
We can read its value from Eq.~\eqref{eq:character_A} by counting the number of constant terms in the sum and dividing by $D_R$, giving us $c'_A = (\nc-1) / (\nc^2-1)$.
This has to be subtracted from $\hat F_R(0) =1$ to get the correct limit.

\bibliographystyle{JHEP-2modlong.bst}
\bibliography{ref}

\end{document}